\documentclass[preprint,5p,times]{elsarticle}
\usepackage{graphics}
\usepackage{epsfig}
\usepackage{amssymb}

\usepackage{tabularx}
\usepackage{subfig}
\usepackage{xfrac}
\usepackage{multirow}
\usepackage{morefloats}

\usepackage{nfontdef}

\newcommand{\ignore}[1]{}

\usepackage{xcolor}

\journal{Computers \& Structures}

\begin{document}

\begin{frontmatter}

\title{Competitive Comparison of Optimal Designs \\ of Experiments for Sampling-based Sensitivity Analysis}

\author[ctu]{Eli\v{s}ka Janouchov\'a}
\ead{eliska.janouchova@fsv.cvut.cz}
\author[ctu]{Anna Ku\v{c}erov\'a\corref{auth}}
\ead{anicka@cml.fsv.cvut.cz}
\cortext[auth]{Corresponding author. Tel.:~+420-2-2435-5326;
fax~+420-2-2431-0775}
\address[ctu]{Department of Mechanics, Faculty of Civil Engineering,
  Czech Technical University in Prague, Th\'{a}kurova 7, 166 29 Prague
  6, Czech Republic}
%

\begin{abstract}
  Nowadays, the numerical models of real-world structures are more
  precise, more complex and, of course, more time-consuming. Despite
  the growth of a~computational performance, the exploration of model
  behaviour remains a complex task. The sensitivity analysis is a
  basic tool for investigating the sensitivity of the model to its
  inputs.  One widely used strategy to assess the sensitivity is based
  on a finite set of simulations for a given sets of input parameters,
  i.e.  points in the design space. An estimate of the sensitivity can
  be then obtained by computing correlations between the input
  parameters and the chosen response of the model. The accuracy of the
  sensitivity prediction depends on the choice of design points called
  the design of experiments. The aim of the presented paper is to
  review and compare available criteria for the assessment of design
  of experiments suitable for sampling-based sensitivity analysis.
\end{abstract}

\begin{keyword}
Design of experiments \sep
Space-filling \sep
Orthogonality \sep
Latin Hypercube Sampling \sep
Sampling-based sensitivity analysis
\end{keyword}

\end{frontmatter}
\section{Introduction}
\label{sec:intro}
Sensitivity analysis (SA) is an important tool for investigating
properties of complex systems. It represents an essential part of
inverse analysis procedures \cite{Kucerova:2007:phd}, response surface
modelling \cite{Helton:2006:RESS2} or uncertainty analysis
\cite{Helton:2006:RESS1}. To be more specific, SA provides some
information about the contributions of individual system
parameters/model inputs to the system response/model outputs. A number
of approaches to SA has been developed, see e.g.  \cite{Saltelli:2000}
for an extensive review. The presented contribution is focused on
widely used sampling-based approaches \cite{Helton:2006:RESS2},
particularly aimed at an evaluation of Spearman's rank correlation
coefficient (SRCC), which is able to reveal a nonlinear monotonic
relationship between the inputs and the corresponding outputs.

When computing the SA in a case of some real system using expensive
experimental measurements or some computationally exhaustive numerical
model, the number of samples to be performed within some reasonable
time is rather limited. Randomly chosen sets of input parameters do
not ensure appropriate estimation of related sensitivities. Therefore
the sets must be chosen carefully. In this contribution we would like
to present a review and comparison of several criteria, which can
govern the stratified generation of input sets -- the so called design of
experiments (DoE).

Generation of optimal DoEs is a very broad topic and all pertinent
aspects cannot be discussed within this paper. Hence, we focus
especially on DoEs in discrete domains.  The presented methods can be
of course applied also to discretised continuous domain.
Nevertheless, other possibilities for generation DoEs in continuous
domains are, however, beyond the scope of this paper.

The following section reviews the criteria for optimisation of DoE,
which are available in literature. Section \ref{LHS} includes some
comments on widely used methods for stratified generation of DoE and
Section \ref{generation} presents the discussion on difficulties
arising from optimisation of particular criteria.
Section~\ref{mutual} is devoted to the comparison of mutual qualities
of particular optimal DoEs and Section~\ref{projection} compares
  their quality in terms of projective properties which are important
  in a screening phase of model analysis. Sequential improvement of
  the existing DoE is discussed in Section \ref{sequential}. Finally,
Sections \ref{sensitivity_fce} and \ref{sensitivity_kce} present the
assessment of the optimal designs quality for usage in sampling-based
SA for theoretical analytical functions and structural models,
respectively.  Concluding remarks are summarised in
Section~\ref{concl}.

\section{Criteria for assessing optimal designs}

A number of different criteria for assessing the quality of particular
DoE can be found in literature. In general, they can be organised into
groups w.r.t. the preferred DoE property. The most widely
preferred features are
\begin{itemize}
\item {\it space-filling} property, which is needed to allow for the
  evaluation of sensitivities valid for the whole given domain of
  admissible input values, the so called design space;
\item {\it orthogonality}, which is necessary to assess the impact of
  individual input parameters.
\end{itemize}
Other main objectives may be preferable in particular applications
of DoE. In response surface methodology, reduction of noise and
bias error can become more important than the orthogonality
\cite{Goel:2008:IJNME}. Nevertheless, no special objectives were
formulated for the case of sampling-based SA, so we employ the common
ones.

\subsection{Space-filling criteria}

Let us recall four widely used space-filling criteria.

\vspace{12pt}

{\bf Audze-Eglais objective function (AE)} proposed by Audze and
Eglais~\cite{Audze:1977} is based on a potential energy among the
design points. The points are distributed as uniformly as possible
when the potential energy $E^{\mathrm{AE}}$ proportional to the
inverse of the squared distances among points is minimised, i.e.
\begin{equation} E^{\mathrm{AE}} = \sum_{i=1}^n \sum_{j=i+1}^n
  \frac{1}{L^2_{ij}} \, ,
  \label{eq:AE}
\end{equation}
where $n$ is the number of the design points and $L_{ij}$ is the
Euclidean distance between points $i$ and $j$.

\vspace{12pt}

{\bf Euclidean maximin (EMM) distance} is probably the
best-known space-filling measure
\cite{Johnson:1990:JSPI,Morris:1995:JSPI}. It states that the
minimal distance $L_{\min,ij}$ between any two points $i$ and $j$
should be maximal. In order to apply the minimisation procedure to
all presented criteria, we minimise the negative value of a minimal
distance $E^{\mathrm{EMM}}$, i.e.
\begin{equation}
  E^{\mathrm{EMM}} = - \min \{..., L_{ij} , ...\}, \quad i = 1...n, \quad j = (i+1) \, ... \, n \, .
  \label{eq:EMM}
\end{equation}

\vspace{12pt}

{\bf Modified $L_2$ discrepancy (ML$_2$)} is a computationally
cheaper variant of a discrepancy measure, which is widely used to assess
precision for multivariate quadrature rules \cite{Fang:1994}.  Here,
the designs are normalised in
each dimension to the interval $\left[ 0, 1 \right]$ and then, the value of ML$_2$ is computed according to
\begin{eqnarray}
  E^{\mathrm{ML}_2} &=& \left( \frac {4}{3} \right) ^k - \frac{2^{(1-k)}}{n} \sum_{d=1}^n \prod_{i=1}^k(3-x_{di}^2) + \nonumber \\
 &+& \frac{1}{n^2} \sum_{d=1}^n \sum_{j=1}^n \prod_{i=1}^k [2-\max(x_{di},x_{ji})] \, ,
  \label{eq:ML}
\end{eqnarray}
where $k$ is the number of input parameters, i.e. the dimension of the
design space and $x_{di}$ and $x_{ji}$ are the $i$-th coordinates of
the $d$-th and $j$-th points, respectively. Since the evaluation of
discrepancy for a large design can be time-consuming, some efficient
algorithms are proposed e.g. in \cite{Heinrich:1996:MC}. To achieve
the best space-filling property of DoE, the value of ML$_2$ should be
minimised.

\vspace{12pt}

{\bf D-optimality criterion (Dopt)} was proposed by Kirsten
Smith in 1918~\cite{Smith:1918} as a~pioneering work in the field of
DoE for regression analysis. This criterion minimises the variance
associated with estimates of regression model coefficients by
minimizing the determinant of the so called dispersion matrix
$(\bf{Z}^T \bf{Z})^{-1}$ or equivalently, by maximizing the
determinant of the so called information matrix $(\bf{Z}^T \bf{Z})$
\cite{Aguiar:1995:CILS}. Again, in order to apply a minimisation
procedure, but to avoid the inversion of the information matrix, we
can minimise negative value of the determinant of the information matrix,
i.e.
\begin{equation}
  E^{\mathrm{Dopt}} = -\det (\bf{Z}^T \bf{Z}) \, ,
  \label{eq:Dopt}
\end{equation}
where $\bf{Z}$ is a matrix with evaluated regression terms in the
design points. In the case of second order polynomial regression and
two-dimensional design space, the matrix becomes
\begin{equation}
  \bf{Z} = \left[\begin{array}{cccccc}
      1&x_{11}&x_{12}&x_{11}^2&x_{12}^2&x_{11}x_{12} \\
      1&x_{21}&x_{22}&x_{21}^2&x_{22}^2&x_{21}x_{22} \\
      \vdots & \vdots & \vdots & \vdots & \vdots & \vdots \\
      1&x_{n1}&x_{n2}&x_{n1}^2&x_{n2}^2&x_{n1}x_{n2}
    \end{array} \right] \, .
  \label{eq:Dopt2}
\end{equation}

\begin{figure*}[h!]
\centering
\includegraphics{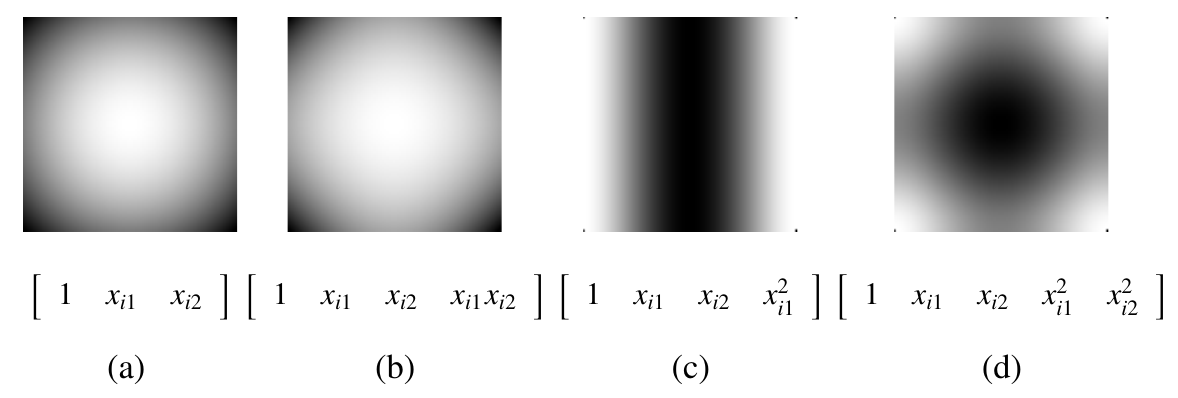}
\caption{Value of D-optimality criterion as a function of fifth
  point's coordinates for different structure of matrix $\bf{Z}$;
  optimal value is represented by black colour.}
\label{fig:Dopt}
\end{figure*}

It is known that under certain conditions, D-optimality criterion
leads to the designs with duplicated points. For illustration, having
five points in two-dimensional space, one can construct only linear
regression having well defined information matrix, i.e. only three
columns in matrix $\bf{Z}$. By fixing the position of four points into
the corners of the squared domain, we can plot the value of
$E^{\mathrm{Dopt}}$ as a function of the fifth point's coordinates.
Figure~\ref{fig:Dopt}a shows that in this situation, the optimal
position is located in one of the occupied corners and the
optimisation will lead to the DoE with duplicates.
To overcome this problem, different approaches have been developed.
For instance, the authors in \cite{Goel:2009:SMO} start with a larger
Latin Hypercube (LH) design optimised w.r.t. EMM criterion.  Then, the
final smaller DoE is selected as a combination of points from the
previously obtained LH design. In this way, no duplicates are
presented, however the optimisation w.r.t.  D-optimal criterion is
quite perfunctory. Another approach based on a Bayesian modification
of an information matrix is proposed in \cite{Hofwing:2010:SMO}. The
idea is to add higher order terms in the response surface
approximation and subsequently to add corresponding columns into the
matrix ${\bf Z}$.  Then, the information matrix $(\bf{Z}^T \bf{Z})$
becomes singular, but according to \cite{DuMouchel:1994:T}, some
constant $\tau \in ]0,1]$ can be added to diagonal elements of
$(\bf{Z}^T \bf{Z})$ corresponding to added columns to overcome this
problem.  The constant $\tau$ is a parameter of the proposed procedure
defining the influence of the Bayesian modification, smaller values
imply smaller influence. It is not obvious which terms should be added
into the matrix ${\bf Z}$, for instance, Figure \ref{fig:Dopt}b shows
that one term needs not be sufficient, and thus this question should
be the subject to additional research. One should keep in mind to add
terms equally to all coordinates so as to preserve the isotropy of the
resulting DoE, see Figure \ref{fig:Dopt}c for an example of
anisotropic criterion. In the case of five points in two-dimensional
domain, two quadratic terms are sufficient to define DoE without the
duplicates, see Figure \ref{fig:Dopt}d. The D-optimal designs
presented further in this paper are obtained by the described manual
Bayesian extension of the information matrix.

\subsection{Orthogonality-based criteria}

There are two well-know approaches to evaluate the orthogonality of a~DoE. The most popular one is based on correlation among the samples'
coordinates, the other one is a condition number.

\vspace{12pt}

{\bf Condition number (CN)} is commonly used in numerical linear
algebra to examine the sensitivities of a linear system
\cite{Cioppa:2007:T}. Here, we use condition number of ${\bf
  X}^T{\bf X}$, where ${\bf X}$ is a matrix of the design points'
coordinates, so called design matrix
\begin{equation}
\bf{X} =  \left[
\begin{array}{cccc}
x_{11} & x_{12} & \cdots & x_{1k} \\
x_{21} & x_{22} & \cdots & x_{2k} \\
\vdots & \vdots & & \vdots \\
x_{n1} & x_{n2} & \cdots & x_{nk} \\
\end{array}
\right]\, ,
\label{eq:doematrix}
\end{equation}
where $n$ is the number of the design points and $k$ is the dimension
of the design space and the columns are centered to sum to $0$ and
scaled to the range $\left[ -1, 1 \right]$. The condition number is
then defined as
\begin{equation}
  E^{\mathrm{CN}} = \mbox{cond}(\bf {X} ^T \bf{X}) = \frac{\lambda_1}{\lambda_n} \, ,
\label{eq:cond}
\end{equation}
where $\lambda_1$ and $\lambda_n$ are the largest and smallest
eigenvalues of ${\bf X}^T{\bf X}$, respectively, therefore the
$E^{\mathrm{CN}}$ is greater or equal to $1$. Values closer to $1$
correspond to more orthogonal DoE, therefore the condition number
should be minimised.

\vspace{12pt}

{\bf Pearson product-moment correlation coefficient (PMCC)} is a~standard measure of a linear dependence between two variables. Having
two variables $x_i$ and $x_j$, the PMCC is defined as
\begin{equation}
  c_{ij} = \frac{\mathrm{Cov}\left(x_i,x_j \right)}{\sigma_{x_i}\sigma_{x_j}} = \frac{\sum_{a=1}^n (x_{ai}-\overline{x_i})(x_{aj}-\overline{x_j})}{\sqrt{\sum_{a=1}^n(x_{ai}-\overline{x_i})^2\sum_{a=1}^n(x_{aj}-\overline{x_j})^2}} \, ,
\label{eq:Pearson}
\end{equation}
where
\begin{equation}
\overline{x_i} = \frac{1}{n}\sum_{a=1}^n x_{ai} \quad \mbox{ and } \quad
\overline{x_j} = \frac{1}{n}\sum_{a=1}^n x_{aj} \, .
\end{equation}
The PMCC takes a value between $-1$ and $1$ and positive values
indicate that the value of $x_i$ tends to increase together with
increasing value of $x_j$, while negative values indicate
decreasing value of $x_i$ with increasing value of $x_j$. Zero
value stands for no linear relationship between $x_i$ and $x_j$.
In order to obtain orthogonal DoE in a multi-dimensional design
space, the PMCC needs to be evaluated for each pair of columns in
the design matrix \eqref{eq:doematrix}. As a result, one obtains
a $k \times k$ symmetric correlation matrix
\begin{equation}
\bf{C} =  \left[
\begin{array}{cccc}
c_{11} & c_{12} & \cdots & c_{1k} \\
c_{21} & c_{22} & \cdots & c_{2k} \\
\vdots & \vdots & & \vdots \\
c_{k1} & c_{k2} & \cdots & c_{kk} \\
\end{array}
\right]\, .
\label{eq:cormatrix}
\end{equation}
In the case of an orthogonal DoE, the correlation matrix ${\bf C}$ is
equal to the identity matrix. To achieve an orthogonal DoE, one can,
for instance, minimise the maximum $|c_{ij}|$ as in
\cite{Cioppa:2007:T} or the sum of squares of the elements above the
main diagonal of ${\bf C}$ as it is done in engineering softwares
\cite{Novak:SPERM,Novak:FREET} as well as in presented results, i.e.
\begin{equation}
E^{\mathrm{PMCC}} = \sqrt{\sum_{i=1}^k \sum_{j=i+1}^k c_{ij}^2} \, .
\label{eq:SPERM2}
\end{equation}

\vspace{12pt}

{\bf Spearman's rank correlation coefficient (SRCC)} can be used
to capture a nonlinear but monotonic relationship between two variables
and therefore, it can be efficiently applied for estimation of
correlations in sampling-based SA
\cite{Helton:2006:RESS2}. The idea is to replace the values of $x_{ai}$
and $x_{aj}$ by their corresponding ranks $r(x_{ai})$ and $r(x_{aj})$
and then the SRCC can be computed as
\begin{equation}
  \rho_{ij} = 1- \frac{6\sum_{a=1}^n \left( r(x_{ai}) - r(x_{aj}) \right)^2}{n(n^2-1)} \, .
\label{eq:Spearman}
\end{equation}
In case of a multi-dimensional design space, the orthogonality of the
DoE can be similarly to \eqref{eq:SPERM2} achieved by minimizing
\begin{equation}
E^{\mathrm{SRCC}} = \sqrt{\sum_{i=1}^k \sum_{j=i+1}^k \rho_{ij}^2} \, .
\label{eq:SPERM2b}
\end{equation}

\vspace{12pt}

{\bf Kendall tau rank correlation coefficient (KRCC)} is an
alternative measure of a nonlinear dependence between two variables.
In particular, it is based on the number of concordant
($T_{c,ij}$) and discordant ($T_{d,ij}$) pairs of samples
according to
\begin{equation}
\tau_{ij} = \frac{T_{c,ij}-T_{d,ij}}{n(n-1)/2} \, ,
\label{eq:Kendall}
\end{equation}
and again, the orthogonal DoE can be obtained by minimizing
\begin{equation}
  E^{\mathrm{KRCC}} = \sqrt{\sum_{i=1}^k \sum_{j=i+1}^k \tau_{ij}^2} \, .
\label{eq:SPERM2c}
\end{equation}

\section{Latin Hypercube Sampling}
\label{LHS}

Since the optimisation of DoE defined on real domains becomes
computationally exhaustive even at moderate number of dimensions or
design points, practical applications are usually restricted to the
optimisation of the so called Latin Hypercube (LH) designs
\cite{Iman:1980}. LH sampling provides a possibility to represent
prescribed probabilistic distribution of particular variables and
hence, it can be efficiently applied in uncertainty analysis
\cite{Helton:2006:RESS1}. The idea is to divide the range of each
variable $x_j$ into $n$ disjoint intervals of equal probability and
one value $x_{ij}$ is selected from each interval. This selection can
be either random or commonly prescribed to the centre of the interval.
Then, the $n$ values for each variable are randomly selected without
replacement and coupled with $n$ values of other variables resulting
in $n$ vectors of variables where each discrete value of each variable
is used only once.

The discretisation itself is quite useful for simplification of the
optimisation process. Therefore, we focus our attention mainly to the
optimisation of DoE in discrete domains assuming that continuous
domains are usually also discretised so as to make the optimisation
process manageable. Of course, the LH restriction simplifies the
optimisation even more, since the search space is significantly
reduced. However, it is not obvious whether such restriction excludes
the best solutions regarding the objective of SA. Moreover, the LH
restriction is not applicable to originally discrete problems, where
design variables are defined in different number of discrete
values/levels and the so called mixed design is needed.  In such a
case, the LH restriction has to be somehow modified. When the numbers
of levels are equal to multiples of each other, one can easily
prescribe the number of points appearing in each level in order to
preserve the homogeneity of resulting designs. In other cases, one can
prescribe only the minimal number of points appearing in each level
\cite{Alam:2004:SMPT,Toropov:2007:ICAAI}.  Alternatively, the authors
in \cite{Vieira:2011:EJOR} optimise the mixed design by mixed integer
programming method. In our numerical experiments, we generate the LH
designs, where the number of points equals the number of levels of one
chosen variable and before the evaluation of a criterion, the
coordinates of points with different number of levels are simply
rounded into the appropriate values. By this way we compare the LH or
the mixed LH (mLH) designs with the free (unrestricted) designs in
order to investigate the impact of the LH restriction to results of
SA.

\section{Optimisation of DoE}
\label{generation}

Efficient generation of optimal DoE is nowadays a subject of an
intensive scientific effort. The main reason is that all of the
presented criteria are multi-modal functions of optimised variables,
hence, gradient-based algorithms cannot be efficiently applied and a
robust stochastic algorithm needs to be developed. Let us recall
several recent works on this topic: the usage of AE criterion for
generating uniform LH designs was recently proposed in
\cite{Bates:2003:AES} and further developed in
\cite{Toropov:2007:ICAAI} applying a permutation-based genetic
algorithm for optimisation; bounds for maximin LH designs were
established in \cite{Dam:2009:OR} and comparison of suitable
generators for the case of EMM criterion are presented in
\cite{Mysakova:2011:EM}; threshold accepting heuristic is applied in
\cite{Fang:2002:MC} for minimizing the ML$_2$ discrepancy on LH
designs; a review of existing methods for generating space-filling
optimal LH designs together with presentation of another heuristic
strategy can be found in \cite{Viana:2010:IJNME} and simulated
annealing was employed for generating LH designs with prescribed
correlation matrix in \cite{Vorechovsky:2009:PEM}. An important
  aspect concerns the sequential generation of DoEs which is briefly
  discussed in Section \ref{sequential}.

Although it is not the aim of the present paper, the efficiency of
a~particular criterion is, indeed, closely related to the possibility
of its optimisation. To sketch the difficulty of generating particular
optimal DoE, we present a~simple comparison of the presented criteria
as functions to be optimised.  For the sake of clarity, we present two
examples of two-dimensional DoEs with four and five points, where
three and four points, respectively, are fixed in the corners of the
squared design space and each criterion is evaluated as a~function of
the last point's coordinates.

\begin{figure*}[h!]
\centering
\includegraphics{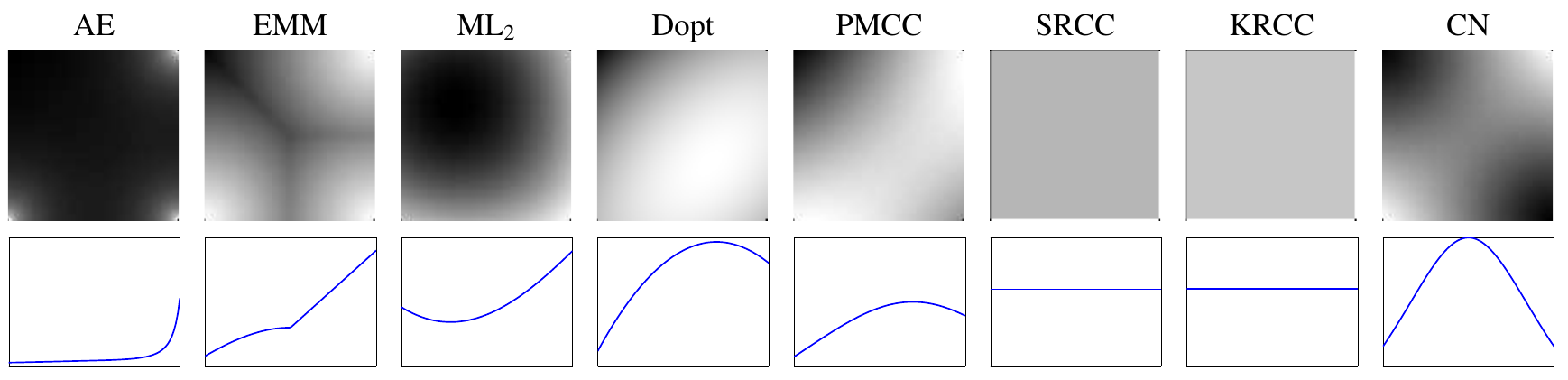}
\caption{Shape of different criteria for varying position of the 4th point.}
\label{fig:bod4}
\end{figure*}

\begin{figure*}[h!]
\centering
\includegraphics{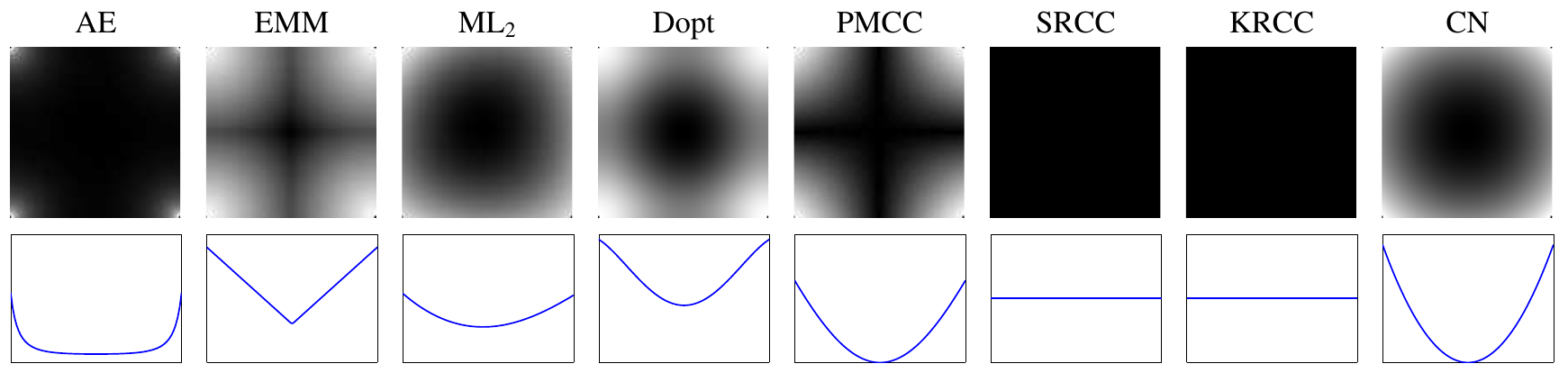}
\caption{Shape of different criteria for varying position of the 5th point.}
\label{fig:bod5}
\end{figure*}

Figures \ref{fig:bod4} and \ref{fig:bod5} show the shape of the
resulting functions in two ways: (i) on the whole domain (the first
row) by intensity of grey colour (black colour represents the minimum)
corresponding to the value of each criterion at the given position of
the last point, and (ii) in more detailed cut of the functional
surface according to the diagonal of the domain (the second row). The
presented results imply several following conclusions concerning an
ease of a corresponding optimisation process.
\begin{itemize}
\item SRCC and KRCC as optimisation criteria are by definition
  discrete functions. When applied to situations with a continuous
  design space, these functions can have constant value on large
  subdomains and this phenomenon can make the optimisation process
  significantly more difficult. Moreover, Figure \ref{fig:bod5} shows
  the situation where the optimal value corresponds to a large
  subdomain of the original space, although the most of randomly
  chosen solutions from this subdomain cannot be intuitively
  considered as suitable for SA.  Therefore, w.r.t. the presented
  example, these two criteria can be considered as the most difficult
  ones to be optimised as well as the criteria poorly defining desired
  DoE.

\item We assume that creation of an excessive number of local extremes
  can be considered as another negative aspect of optimised
  criterion.  Figure \ref{fig:bod4} demonstrates that criteria PMCC,
  CN and Dopt have undesirable local extreme in one of the occupied
  corner of the design space pointing at the inconvenient character of
  these criteria. Moreover, the CN criterion evaluates the design with
  duplicated points in the bottom-right corner to be as good as the
  design with single point in each corner. That is definitely
  undesirable, because the design with duplicates is neither
  orthogonal nor space-filling.

\item Other, but less inconvenient, feature of optimised functions is
  non-smoothness appearing in the case of EMM criterion in both
  Figures~\ref{fig:bod4} and \ref{fig:bod5}. It was already mentioned
  that in optimisation of the whole DoE, the all presented criteria
  become multi-modal and gradient-based methods cannot be efficiently
  applied.  Therefore, it is questionable, how much is the smoothness
  important when stochastic optimisation methods such as simulated
  annealing or evolutionary algorithms are employed to solve this
  problem.

\item The last feature arising from the presented examples concerns
  the ML$_2$ criterion. From Figure \ref{fig:bod4} it is clearly
  visible that the optimal position of the fourth point w.r.t. this
  criterion is not in the free corner but rather inside the free
  quadrant of the domain. Such an optimal design is slightly worse in
  terms of space-filling as well as orthogonality, but it remains the
  subject of other tests to evaluate the quality of ML$_2$ criterion
  w.r.t. usage in SA.

\end{itemize}

To conclude this section, we would like to point out that the AE
criterion seems to have best properties regarding the subsequent
optimisation process.

Following sections presents comparisons of DoEs optimised w.r.t.
described criteria. Since the designs are not excessively complex, the
Simulated Annealing method \cite{Kirkpatrick:1983:S,Cerny:1985:JOTA}
was applied to optimise each criterion. The procedure slightly differs
for the free and for the LH designs. In the first case a single loop
of the algorithm involves a sequential selection of a design point and
its random movement to any unoccupied position. In the latter case the
algorithm randomly chooses two points and then switches one of their
randomly chosen coordinates, see \cite{Vorechovsky:2009:PEM} for more
detailed description of the algorithm implementation for correlation
control in small-sample LH designs. The acceptance of a new solution
is driven by the Metropolis criterion
\begin{equation}
  \exp \left( \frac{f_{\mathrm{old}} - f_{\mathrm{new}}}{T} \right) \geq U \, ,
\end{equation}
where $f_{\mathrm{old}}$ and $f_{\mathrm{new}}$ stand for values of
a~criterion for an actual and for a~new solution, respectively. $T$
denotes the algorithmic temperature initially set to $T_{\mathrm{max}}
= 10^{-3}$ and gradually reduced by a multiplicative constant
$T_{\mathrm{mlt}} =
(\sfrac{T_{\mathrm{max}}}{10^{-6}})^{\sfrac{1}{100}}$ after each
$\sfrac{n_{\mathrm{max}}}{10}$ iterations or sooner if the number of
accepted movements reaches the value $\sfrac{n_{\mathrm{max}}}{100}$.
The entire algorithm terminates after $n_{\mathrm{max}} = 10^6$
objective function evaluations.

Of course, there is no guarantee that the global optimum is achieved,
nevertheless, more frequent convergence to local extremes also reflect
the shortcoming of a particular criterion. Hence, we decided to
present the obtained results without any deeper search for more robust
and reliable optimisation method.

\section{Comparison of mutual qualities of optimal DoEs}
\label{mutual}

While the presented criteria were originally developed to achieve good
space-filling property or orthogonality of the final DoE, it does not
mean that the resulting optimal DoE must have bad quality w.r.t. other
properties. In order to estimate the mutual quality of individual
criteria, we have performed a set of tournament comparisons with small
designs having $7$, $10$ and $13$ points in two-dimensional discrete
domain. The first variable can only attain $10$ discrete values, while
the second variable has possible $7$, $10$ or $13$ values according
to the number of design points.

\begin{figure*}[h!]
\centering
\includegraphics{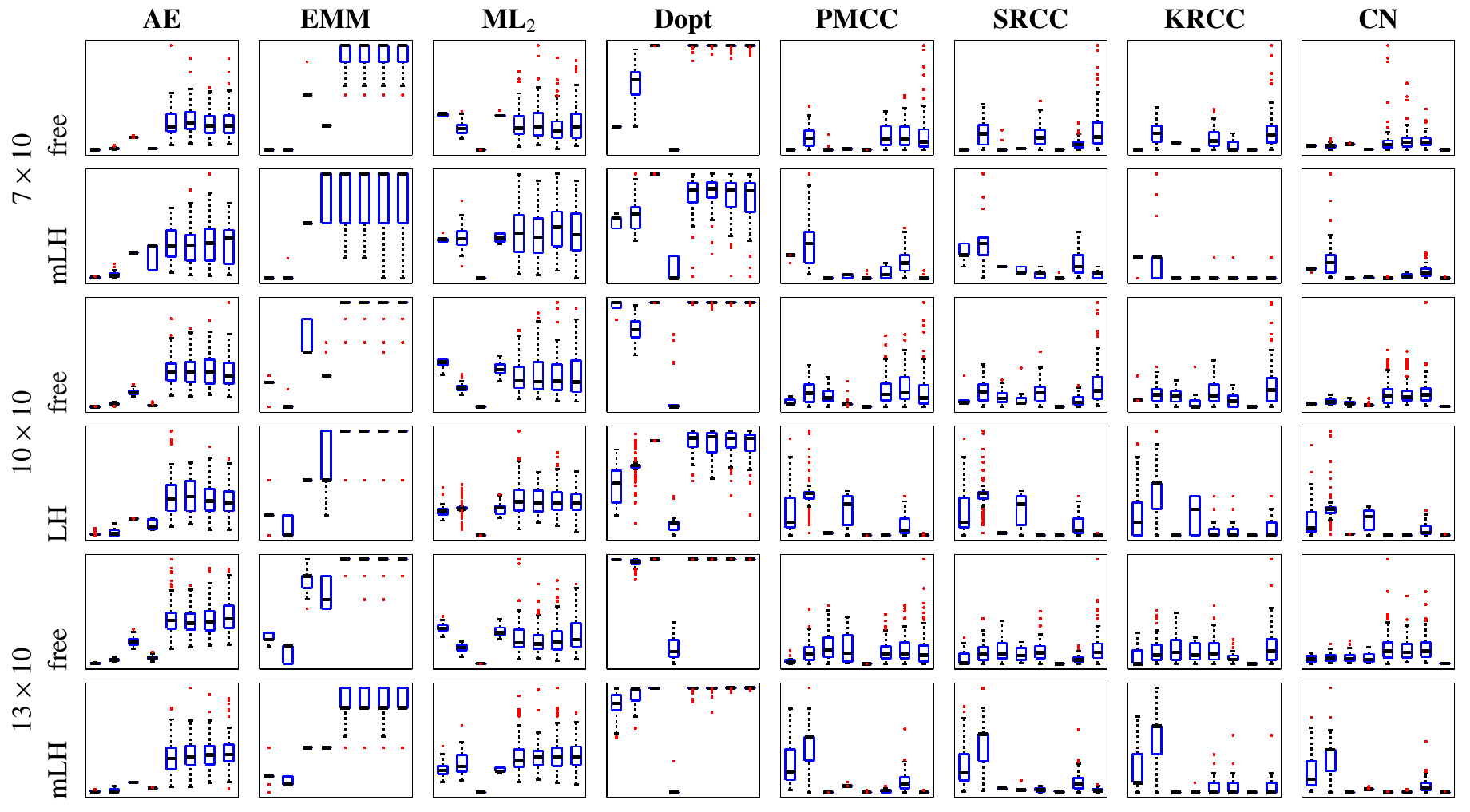}
\caption{Statistics on results of criteria mutual comparison. Each
  rectangle contains eight box plots representing the distribution of
  the values of a~criterion associated with the corresponding column.
  Therefore the scale of each rectangle changes according to the
  evaluated criterion. Individual box plots correspond to results
  obtained by optimizing particular criteria. They are ordered in the
  same way as columns: AE, EMM, ML$_2$, Dopt, PMCC, SRCC, KRCC and CN.
  The rows correspond to the size of the design space and the type of
  a restriction applied on the designs.}
\label{fig:mutual}
\end{figure*}

\begin{figure*}[h!]
\centering
\includegraphics{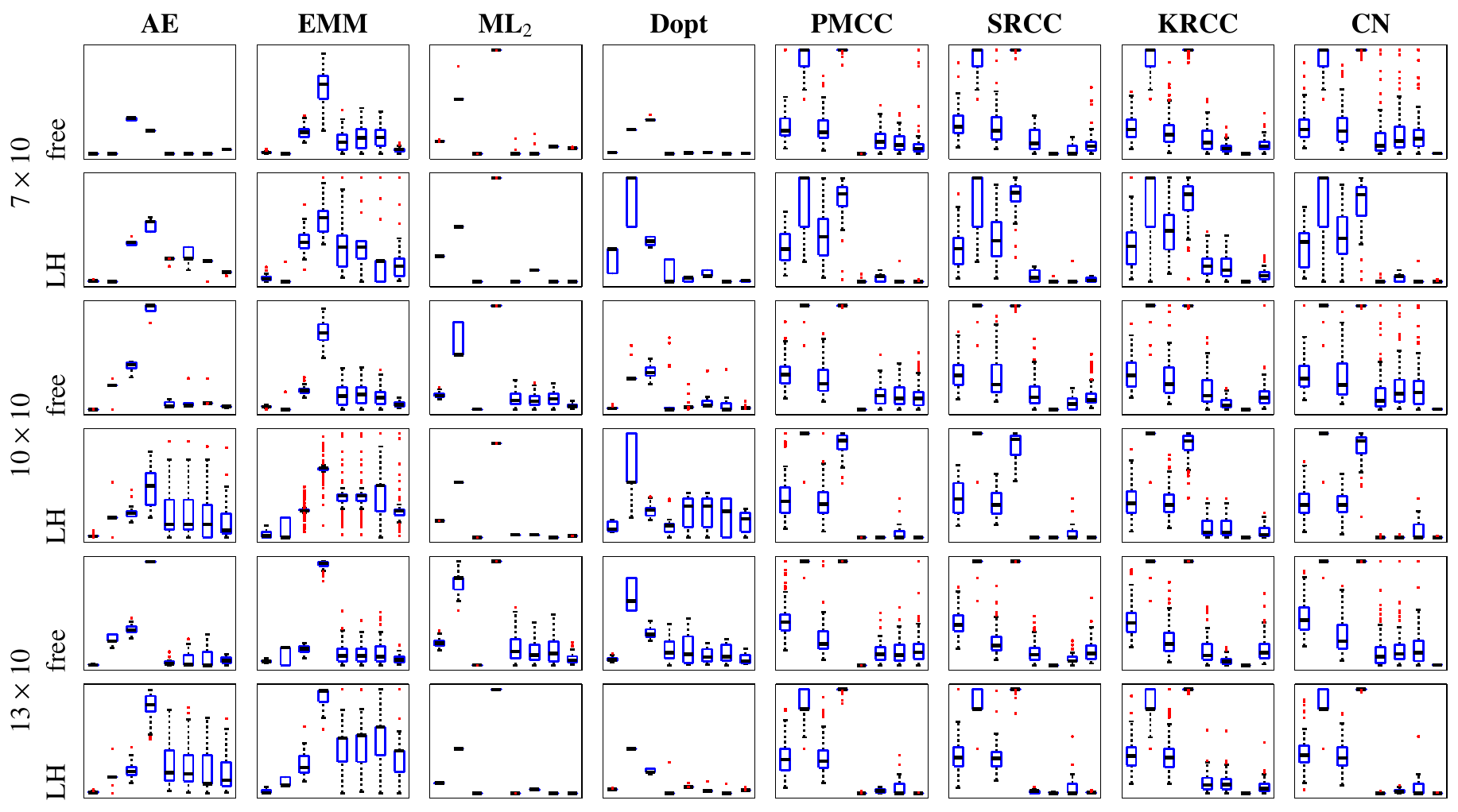}
\caption{Statistics on results of criteria mutual comparison organised
  in different way than in Figure \ref{fig:mutual}. Each rectangle
  contains eight box plots, all corresponding to an optimised
  criterion associated with the corresponding column. Values in
  individual box plots are obtained by evaluating particular criteria,
  ordered in the same way as columns: AE, EMM, ML$_2$, Dopt, PMCC,
  SRCC, KRCC and CN. The scale of each box plot is retained the same
  as in Figure \ref{fig:mutual}. The rows correspond again to the size
  of the design space and the type of a restriction applied on the
  designs.}
\label{fig:mutual2}
\end{figure*}

In order to reduce the effect of randomness of the optimisation
algorithm, the optimisation process was performed $100$ times for each
criterion and the obtained designs were stored and subsequently
evaluated by all other criteria. Figures \ref{fig:mutual} and
\ref{fig:mutual2} show statistics over the obtained results in terms
of box plots. In Figure \ref{fig:mutual}, each rectangle is devoted to
one particular criterion, which is used to evaluate the DoEs obtained
by optimizing itself and all other criteria, while in Figure
\ref{fig:mutual2}, each rectangle contains results of DoEs optimised
w.r.t. one criterion and then evaluated by all the other criteria.
Repeating the results in Figures \ref{fig:mutual} and
\ref{fig:mutual2} in different arrangement gives us the possibility to
more easily formulate distinct conclusions. While the Figure
\ref{fig:mutual} demonstrates the ease of satisfaction of each
criterion, the Figure \ref{fig:mutual2} shows the quality of
particular optimal DoEs when evaluated by other criteria. Just recall
that all the criteria are minimised and smaller values represent
better designs.

Regarding the results plotted in Figures \ref{fig:mutual} and
\ref{fig:mutual2}, the conclusions can be stated as follows:

\begin{itemize}
\item The optimal values obtained by optimisation of particular
    criterion in all of the $100$ runs have very small scatter, which
    indicates that the global optimum was probably found in most of
    the cases. It means that Simulated Annealing method is robust
    enough to solve these optimisation problems in a given time.  Thus
    the results should remain the same even if some other sufficiently
    robust algorithm is applied.
\item The nonorthogonality is easier to be minimised, which is a
    conclusion consistent with the results derived in
    \cite{Vorechovsky:2012:PEM}. The overall results show that the
  designs optimised in terms of space-filling property are often
  nearly orthogonal. On the other hand, the designs optimised w.r.t.
  the orthogonality display usually bad space-filling and the
  improvement achieved by LH restriction is negligible. What is more
  surprising is that the resulting designs are not so well evaluated
  even by other orthogonality-based criteria.
\item The AE and EMM optimal designs exhibit similar properties and
  good qualities w.r.t. each other. However their quality is
  not very good in terms of ML$_2$ and Dopt criteria. The AE optimal
  designs are slightly more non-orthogonal. The orthogonality of both
  is deteriorated by applying LH restrictions.
\item Dopt criterion is a unique criterion in the sense that all
  designs optimised w.r.t. all other criteria are very far from being
  D-optimal. On the other hand, the D-optimal designs are very good
  regarding the AE criterion and moderate in EMM and ML$_2$ criteria.
  These qualities are slightly deteriorated by applying LH
  restrictions. Regarding the orthogonality of the D-optimal designs,
  LH restrictions have unclear effect. In general, the D-optimal
  designs have good or very good level of orthogonality. In the case
  of 10 design points, LH restriction leads to worsening of the
  orthogonality, but it leads to an improvement in the case of 13
  design points.
\item ML$_2$ optimal designs are slightly worse in terms of AE and EMM
  criteria then D-optimal designs, but they have good level of
  orthogonality which is even improved in combination with the LH
  restriction.
\end{itemize}

\begin{figure*}[h!]
\centering
\includegraphics{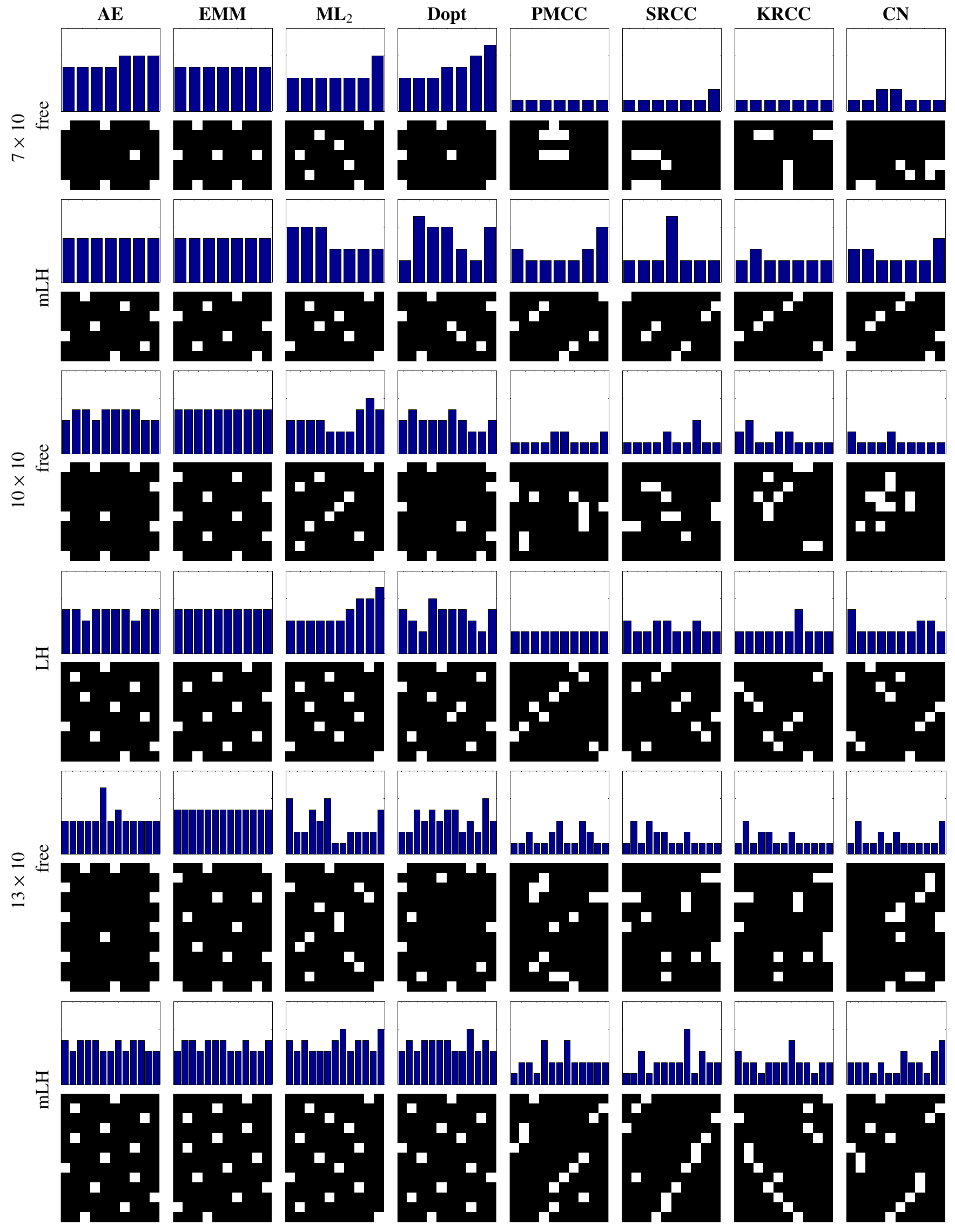}
\caption{Optimal designs and corresponding bar charts of minimum distances.}
\label{fig:plots}
\end{figure*}

To appreciate the space-filling properties more visually, the
resulting optimal DoEs are plotted in Figure \ref{fig:plots} together
with the corresponding bar charts visualising the distance of each
point to its nearest neighbour. The plotted examples are chosen among
the other designs for their worst result in the sum of the minimal
distances to other points. The aim is to show the worst results one
can get by optimizing particular criteria.

\section{Projective properties of optimal DoEs}
\label{projection}

Besides the space-filling and non-orthogonality property, the quality
in terms the projective properties is the most crucial in sensitivity
analysis. A DoE has good projective properties if each coordinate of
each design point is strictly unique
\cite{Bursztyn:2006,Crombecq:2011:EJOR}. A SA -- in the so called
screening phase -- can often reveal model parameters with a negligible
impact on model response. These parameters are then omitted in the
following analysis and further application of the created DoE such as
in response surface construction. Omitting of some parameters originally
involved in constructed DoE implies its projection from $k$-dimensinal
space onto $(k-u)$-dimensional space, where $k$ stands for the
original number of parameters and $u$ is the number of unimportant
parameters to be further neglected. If none of the design points is
projected onto another point, the DoE has good projective properties
which are alternatively called as non-collapsing property
\cite{Dam:2007:OR}.  In an opposite situation, the projection results
in smaller DoE with duplicate points and the corresponding simulations
are wasted.

\begin{figure*}[h!]
\centering
\includegraphics{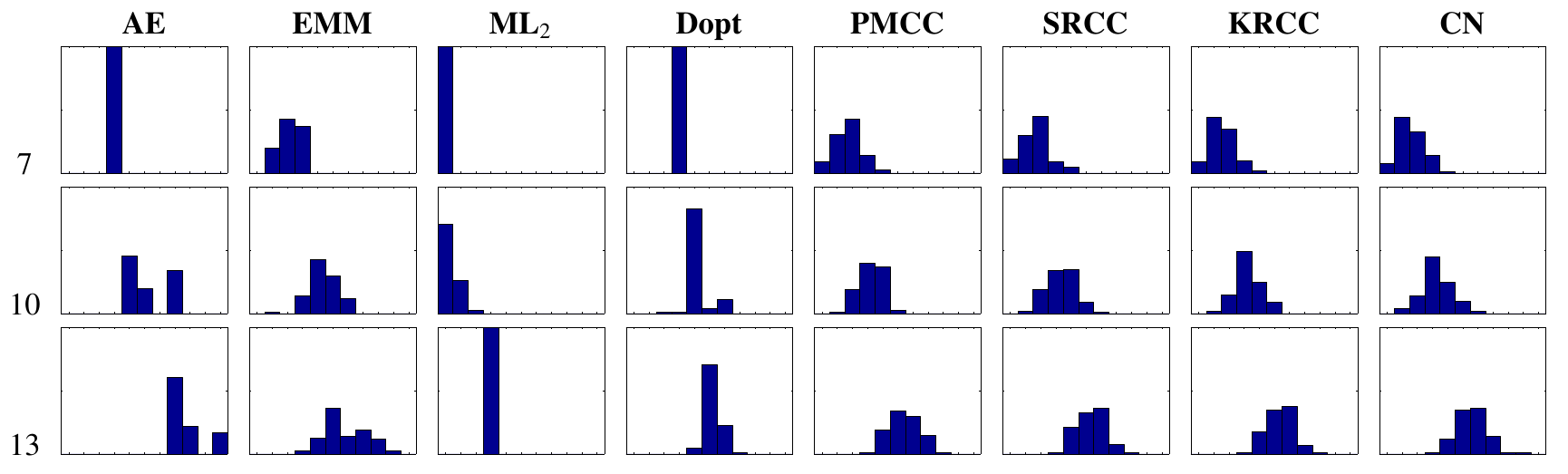}
\caption{Histograms of redundant points in projected DoEs. The
  horizontal axis corresponds to the number of redundant points from
  $0$ to $10$, while the vertical axis represents the number of DoEs
  with the given number of redundant points from $0$ to $100$. Rows
  correspond to the number of levels of the variable left out. Columns
  correspond again to particular criteria.}
\label{fig:projective_hist}
\end{figure*}

An important advantage of Latin Hypercube Sampling (LHS) is that the
resulting DoEs have good projective properties given by definition of
the LHS itself. Hence, we were interested in a projective properties
of free designs optimised w.r.t. the particular criteria. For
measuring the quality in terms of projective properties, the authors
in \cite{Bursztyn:2006} use the so called minimum projected distance,
which is suitable for DoEs defined in continuous domains. For case of
discrete domains, we use simply the number of redundant points. To
that purpose, we have projected all the optimal designs presented in
the previous section onto the dimension discretised into $10$ values.
The obtained results are plotted in Figure \ref{fig:projective_hist}
in terms of histograms representing the number of DoEs with the given
number of redundant points.

For an easier interpretation of the results, the average number of
redundant points are listed in Table \ref{tab:projective_mean}.
\begin{table*}[h!]
\centering
\begin{tabular}{ccccccccc}
& {\bfseries AE} & {\bfseries EMM} & {\bfseries ML$_2$} & {\bfseries Dopt} & {\bfseries PMCC} & {\bfseries SRCC} & {\bfseries KRCC} & {\bfseries CN}\\
\hline
\hline
$7$ &
\footnotesize 3,00 & \footnotesize 2,17 & \footnotesize 0,00  & \footnotesize 3,00 &
\footnotesize 1,71  & \footnotesize 1,67 & \footnotesize 1,52  & \footnotesize 1,56
\\
$10$ &
\footnotesize 5,22 & \footnotesize 4,37 & \footnotesize 0,32  & \footnotesize 4,23 &
\footnotesize 3,22  & \footnotesize 3,33 & \footnotesize 3,24  & \footnotesize 3,29
\\
$13$ &
\footnotesize 7,73 & \footnotesize 5,81 & \footnotesize 3,00  & \footnotesize 5,20 &
\footnotesize 5,42  & \footnotesize 5,32 & \footnotesize 5,35  & \footnotesize 5,57
\\
\hline
celkem &
\footnotesize 5,32 & \footnotesize 4,12 & \footnotesize 1,11  & \footnotesize 4,14 &
\footnotesize 3,45  & \footnotesize 3,44 & \footnotesize 3,37  & \footnotesize 3,47
\end{tabular}
\caption{Average number of redundant simulations.}
\label{tab:projective_mean}
\end{table*}
The results show the superiority of ML$_2$ designs containing minimal
number of redundant points in projected designs. Worst results were
obtained for other space-filling designs.

\section{Sequential generation of DoEs}
\label{sequential}

Before generation of a desired DoE, one has to make a decision about
the number of design points. In case of a discrete problem, the
decision usually follows the number of variable levels, while in
continuous problems, the only clue is usually the available time for
running the simulations. The authors in \cite{Crombecq:2011:EJOR}
consider so called granularity as an important property of a DoE
generation strategy. A fine-grained sequential strategy adds just a
few points in each iteration, while the coarse-grained adds new points
in large batches. Methods allowing only one iteration produce so
called one-shot DoEs. The principal advantage of fine-grained
sequential strategy is the possibility to avoid the generation of too
many samples which are not necessary (i.e. oversampling) or too few
samples which are not sufficient to achieve the desired accuracy (i.e.
undersampling). On the other hand, the advantage of one-shot DoEs is
generally their better quality in terms of space-filling or
non-orthogonality and of course in terms of any chosen criterion.
However, the optimisation of a large design can be more complex and
the obtained locally optimal design may display worse properties then
coarse-grained sequentially produced design.

The authors in \cite{Crombecq:2011:EJOR} present a review of a
different procedures for one by one generated sequential designs and
compare their qualities with one-shot DoEs in terms of space-filling
and non-orthogonality property. The presented methods are searching
new points position in continuous domain or in a sequentially refined
grid in order to provide a Latin Hypercube-like designs. The latter
idea is elaborated and tested in more detail e.g. in
\cite{Vorechovsky:2009:Topping}. All these methods are stochastic and in
each iteration, some sort of optimisation process is applied in order
to increase the space-filling or non-orthogonality of the DoE.

Another type of methods are purely deterministic, such as factorial
designs \cite{Montgomery:2001} or methods used for numerical
integration such as sparse grids, see e.g. \cite{Smolyak:1963,
  Keese:2003:TR}. The advantage of these pre-optimised sequential
designs is very fast generation free of any optimisation at every
iteration. The cost is of course generally worse space-filling or
non-orthogonality.

Last group of methods we should mention consists of strategies
employed in response surface modelling, where the new sample points
are placed where the response surface is supposed to have large error,
see e.g.  \cite{Jones:1998:JGO} or \cite{Lehman:2004}.  However, in
sensitivity analysis we cannot use any such additional information.
Moreover, in sensitivity analysis it is also much harder to estimate
the accuracy of the sensitivity predictions and decide whether the DoE
is large enough. One way is to compute the bootstrap confidence
intervals, see e.g. \cite{Efron:1993}.

In this paper, we do not presume to discuss this broad topic in more
details and we focus only on sequential designs in discrete domains,
where the grid is fixed. Therefore, we employed two strategies, where
the number of new points is equal to the number of levels in discrete
domain and remains the same in all iterations. At each iteration, the
positions of new points are optimised w.r.t. the chosen criterion,
which is always evaluated for the whole design. The designs presented
in Section \ref{generation} are used as a starting DoEs and the
sequential strategy is applied to each of $100$ DoEs in order to limit
again the randomness of the optimisation algorithm.

\begin{figure*}[h!]
\centering
\includegraphics{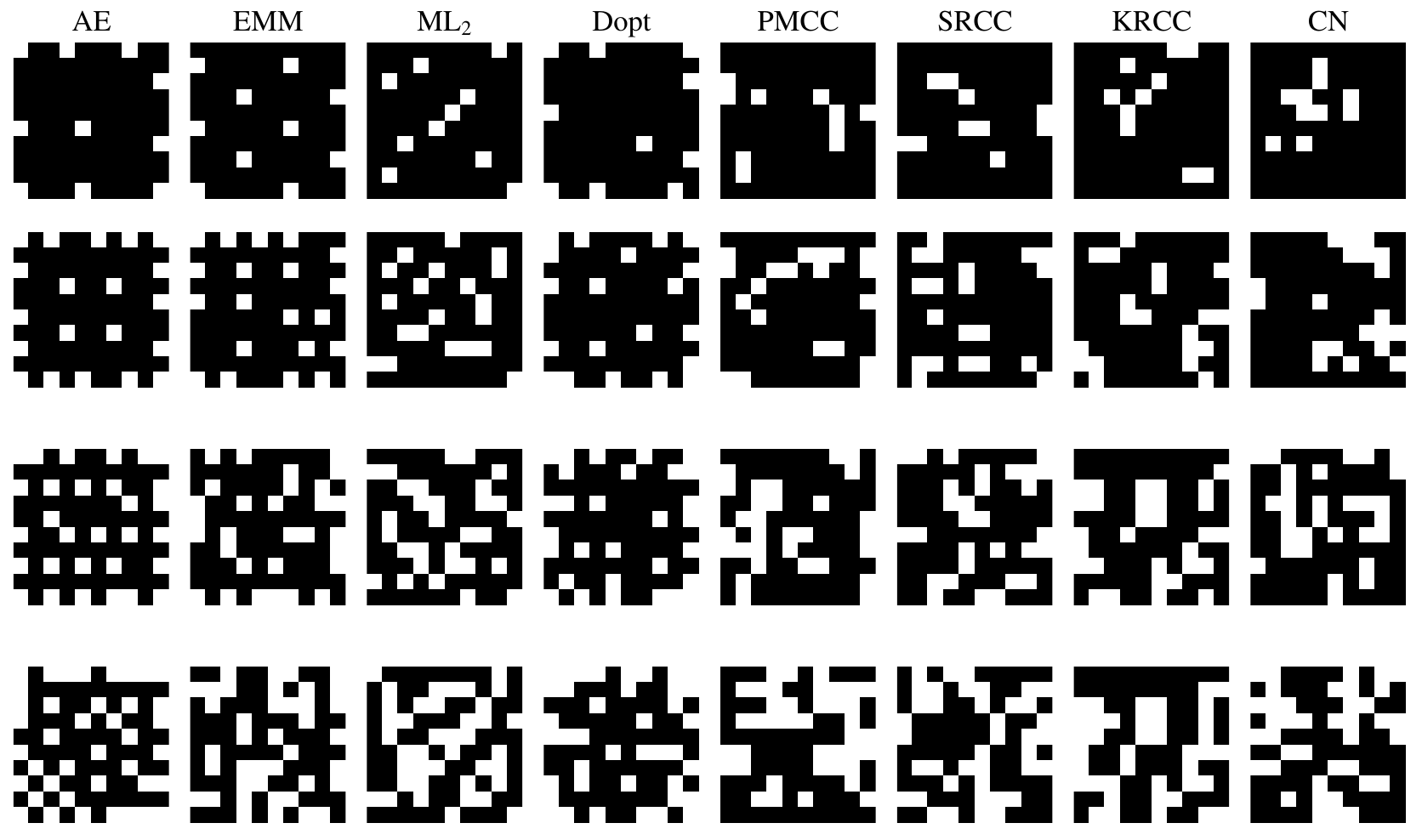}
\caption{Sequential free designs.}
\label{fig:seqfree}
\end{figure*}

The first strategy is based on free selection of new points according
to the optimized criterion and the free DoEs were employed as initial
ones, see Figures \ref{fig:seqfree}. The plotted examples are chosen
among the other designs for their worst result in the sum of the
minimal distances to other points. The aim is to show the worst
results one can obtain by optimizing particular criteria.

\begin{figure*}[h!]
\centering
\includegraphics{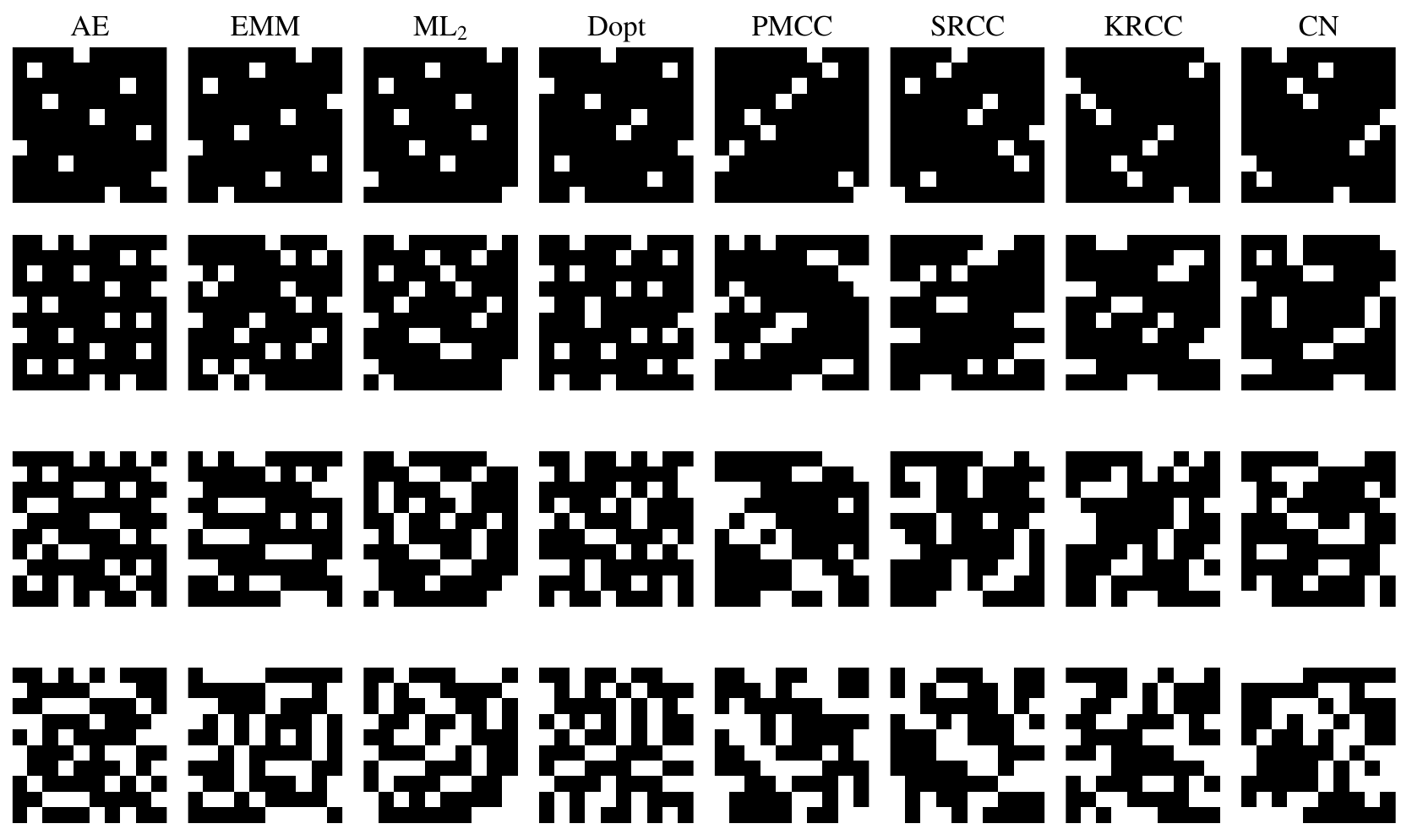}
\caption{Sequential LH designs.}
\label{fig:2seqLH}
\end{figure*}

The second strategy starts from the LH designs and preserves the LH
constrains also for the added points and thus, the equal number of
points are located in each column or row. The~resulting DoEs obtained
by this method are presented in Fig.  \ref{fig:2seqLH}.

The quality of the presented designs is examined directly within the
sensitivity analysis presented in the following section.

\section{Sensitivity analysis on a set of analytical functions}
\label{sensitivity_fce}

Although the SRCC criterion achieved bad results for generating
optimal DoE, it has been shown that it can significantly improve the
results in estimating the importance of model parameters in
sensitivity analysis for the case of nonlinear monotonic
models~\cite{Helton:2006:RESS2}.

Having the numerical model given as
\begin{equation}
  z = f(x_1, x_2, \dots, x_k)
\label{eq:model}
\end{equation}
relating the model response $z$ and the model parameters $x_i$, the
impact of the parameter $x_i$ to the model response $z$ can be
estimated by evaluating their Spearman's rank correlation
$\rho_{x_i,z}$ according to
\begin{equation}
  \rho_{x_i,z} = 1- \frac{6\sum_{a=1}^n \left( r(x_{ai}) - r(z_{a}) \right)^2}{n(n^2-1)} \, ,
\label{eq:Spearman2}
\end{equation}
where $x_{ai}$ are values of particular model parameter corresponding
to points in DoE and $z_a$ are values of model responses corresponding
to these points.

\begin{figure*}[h!]
\centering
\includegraphics{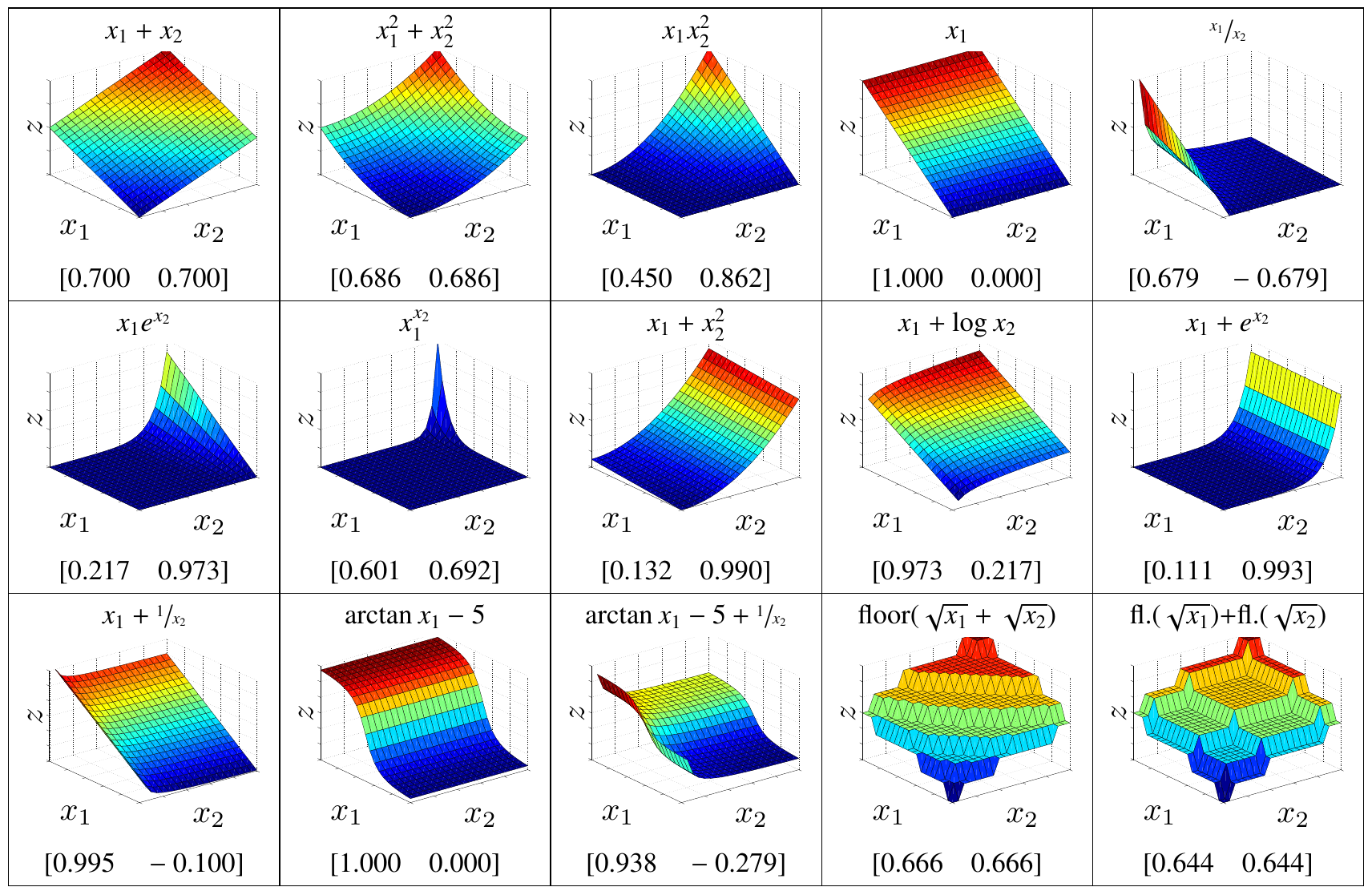}
\caption{Shapes of 15 functions for sensitivity analysis with the
  corresponding values of parameter-response correlations.}
\label{fig:shapes}
\end{figure*}

In engineering practice, the majority of the numerical models fulfil
the condition of a monotonic relationship between the model parameters
and the model response. Therefore, to support the study of optimal DoE
quality in sampling-based SA, we performed the same comparison for a
list of nonlinear but monotonic models. In particular, we consider the
two-parametric models with discrete parameters, one with $10$ feasible
discrete values and the second attaining $7$, $10$ or $13$ feasible
discrete values according to the employed DoE. The shapes of the
chosen models plotted for the case of square $10 \times 10$ domain are
shown in Figure \ref{fig:shapes} together with corresponding
parameter-response correlations obtained for the {\bf Full} design
consisting of all feasible design points (here $100$ points).

\begin{figure*}[h!]
\centering
\includegraphics{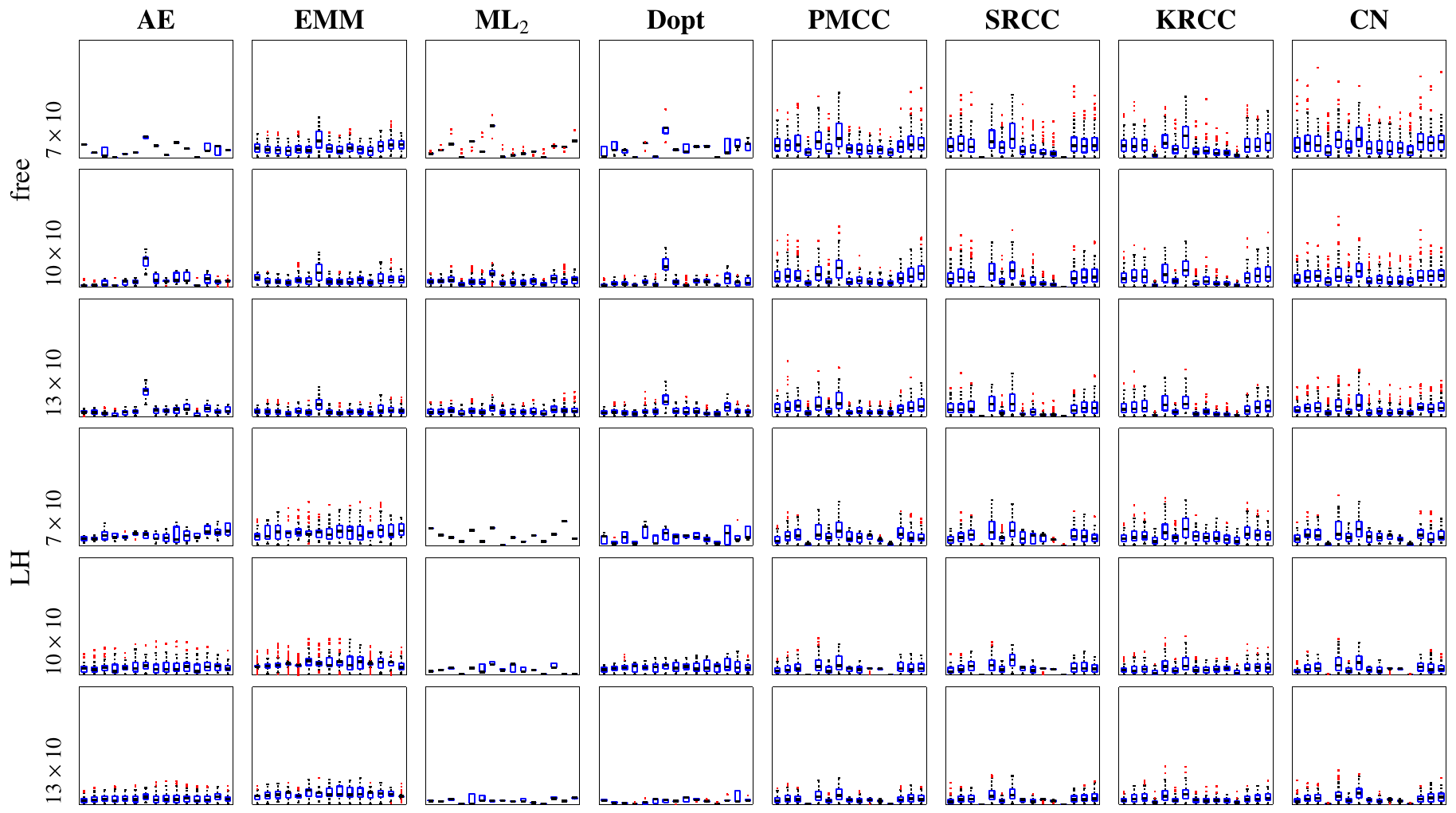}
\caption{Statistics on results of criteria in estimating
  parameter-response correlations for analytical models. Each
  rectangle contains $15$ box plots representing the distribution of
  errors in correlation prediction for particular models depicted in
  Figure \ref{fig:shapes}. The scale is $0$ at the bottom and $1$ at
  the top per rectangle. Furthermore, each rectangle refers to the
  optimal designs according to one criterion associated with the
  corresponding column and to the chosen size of the design space and
  the type of restriction applied associated with the corresponding
  row.}
\label{fig:citl_fce}
\end{figure*}

\begin{figure*}[h!]
\centering
\includegraphics{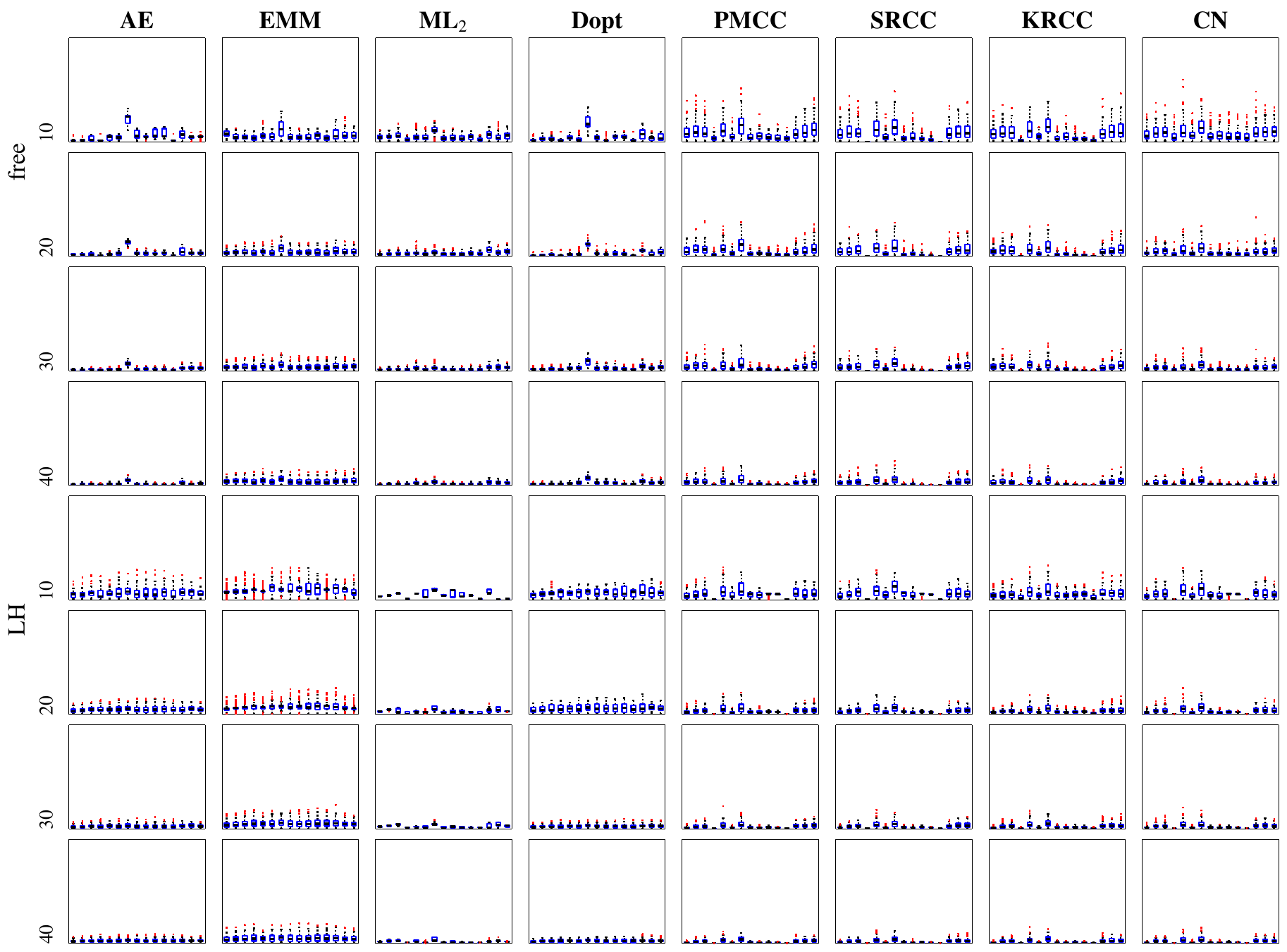}
\caption{Statistics on sequential DoEs in estimating
  parameter-response correlations for analytical models. Each
  rectangle contains $15$ box plots representing the distribution of
  errors in correlation prediction for particular models depicted in
  Figure \ref{fig:shapes}. The scale is $0$ at the bottom and $1$ at
  the top per rectangle. Furthermore, each rectangle refers to the
  optimal designs according to one criterion associated with
  the~corresponding column and to the chosen number of the design points and
  the type of restriction applied associated with the corresponding
  row.}
\label{fig:citl_fce_seq}
\end{figure*}

Then, the parameter-response correlations were estimated using the all
optimal designs involved in the mutual comparison in the previous
section and the differences among correlations $\tilde{\rho}$ obtained
by the optimal designs and correlations $\rho$ obtained by the full
designs are stored. The error measure $\epsilon$ in the
parameter-response correlations evaluated for a given function is
considered as an average difference between each parameter and model
response correlation obtained by an optimal and a full design, i.e.
\begin{equation}
  \epsilon = \frac{1}{k} \sum_{i=1}^{k} | \tilde{\rho}_{x_i,z} - \rho_{x_i,z} | \, .
\label{eq:error}
\end{equation}
The statistics over the obtained values of errors $\epsilon$ is
presented in Figures~\ref{fig:citl_fce} and~\ref{fig:citl_fce_seq}
again using the box plots.

For an easier evaluation of particular criteria, the mean and maximal
errors over all models multiplied by $100$ are listed in Tables
\ref{tab:citl_fce} and \ref{tab:citl_fce_seq}.
\begin{table*}[h!]
\centering
\tabcolsep=2pt
\begin{tabular}{cc|cc|cc|cc|cc|cc|cc|cc|cc}
& & \multicolumn{2}{c|} {\bfseries AE} & \multicolumn{2}{c|} {\bfseries EMM} & \multicolumn{2}{c|} {\bfseries ML$_2$} & \multicolumn{2}{c|} {\bfseries Dopt} & \multicolumn{2}{c|} {\bfseries PMCC} & \multicolumn{2}{c|} {\bfseries SRCC} & \multicolumn{2}{c|} {\bfseries KRCC} & \multicolumn{2}{c} {\bfseries CN}\\
  &        & \footnotesize mean & \footnotesize max & \footnotesize mean & \footnotesize max & \footnotesize mean & \footnotesize max & \footnotesize mean & \footnotesize max & \footnotesize mean & \footnotesize max & \footnotesize mean & \footnotesize max & \footnotesize mean & \footnotesize max & \footnotesize mean & \footnotesize max\\
\hline
\hline
\multirow{2}{*}{\rotatebox{90}{free}} & $7 \times 10$ &
\footnotesize 6.7  & \footnotesize 18.0 &
\footnotesize 8.5  & \footnotesize 34.5 &
\footnotesize 8.4  & \footnotesize 36.3 &
\footnotesize 7.8  & \footnotesize 41.2 &
\footnotesize 10.5 & \footnotesize 59.3 &
\footnotesize 9.7  & \footnotesize 60.8 &
\footnotesize 9.4  & \footnotesize 51.8 &
\footnotesize 12.2 & \footnotesize 76.6
\\
& $10 \times 10$ &
\footnotesize 5.9 & \footnotesize 32.0 &
\footnotesize 6.2 & \footnotesize 29.4 &
\footnotesize 5.1 & \footnotesize 20.5 &
\footnotesize 4.5 & \footnotesize 33.6 &
\footnotesize 8.4 & \footnotesize 51.4 &
\footnotesize 7.3 & \footnotesize 48.5 &
\footnotesize 7.5 & \footnotesize 46.5 &
\footnotesize 8.6 & \footnotesize 59.9
\\
& $13 \times 10$ &
\footnotesize 5.3 & \footnotesize 30.8 &
\footnotesize 4.7 & \footnotesize 24.9 &
\footnotesize 4.6 & \footnotesize 20.8 &
\footnotesize 4.8 & \footnotesize 29.7 &
\footnotesize 6.8 & \footnotesize 47.1 &
\footnotesize 5.6 & \footnotesize 37.0 &
\footnotesize 6.1 & \footnotesize 39.9 &
\footnotesize 6.9 & \footnotesize 39.6
\\
\cline{2-18}
& overall&
\footnotesize \bf 6.0 & \footnotesize \bf 32.0 &
\footnotesize \bf 6.5 & \footnotesize \bf 34.5 &
\footnotesize \bf 6.0 & \footnotesize \bf 36.3 &
\footnotesize \bf 5.7 & \footnotesize \bf 41.2 &
\footnotesize \bf 8.6 & \footnotesize \bf 59.3 &
\footnotesize \bf 7.5 & \footnotesize \bf 60.8 &
\footnotesize \bf 7.7 & \footnotesize \bf 51.8 &
\footnotesize \bf 9.2 & \footnotesize \bf 76.6
\\
\hline
\hline
\multirow{2}{*}{\rotatebox{90}{LH}} & $7 \times 10$ &
\footnotesize 9.0  & \footnotesize 20.6 &
\footnotesize 11.9 & \footnotesize 37.3 &
\footnotesize 8.5  & \footnotesize 20.8 &
\footnotesize 7.0  & \footnotesize 21.9 &
\footnotesize 7.6  & \footnotesize 37.3 &
\footnotesize 7.4  & \footnotesize 39.1 &
\footnotesize 8.8  & \footnotesize 40.5 &
\footnotesize 7.9  & \footnotesize 42.9
\\
& $10 \times 10$ &
\footnotesize 6.7 & \footnotesize 28.9 &
\footnotesize 9.8 & \footnotesize 31.1 &
\footnotesize 4.5 & \footnotesize 11.2 &
\footnotesize 6.8 & \footnotesize 20.3 &
\footnotesize 5.5 & \footnotesize 31.4 &
\footnotesize 5.8 & \footnotesize 28.4 &
\footnotesize 5.6 & \footnotesize 33.0 &
\footnotesize 5.3 & \footnotesize 30.7
\\
& $13 \times 10$ &
\footnotesize 5.2 & \footnotesize 20.0 &
\footnotesize 8.7 & \footnotesize 22.5 &
\footnotesize 3.3 & \footnotesize 8.9 &
\footnotesize 2.8 & \footnotesize 11.6 &
\footnotesize 4.4 & \footnotesize 22.7 &
\footnotesize 4.3 & \footnotesize 24.9 &
\footnotesize 4.9 & \footnotesize 32.2 &
\footnotesize 4.6 & \footnotesize 29.1
\\
\cline{2-18}
& overall &
\footnotesize \bf 7.0 &  \footnotesize \bf 28.9 &
\footnotesize \bf 10.1 & \footnotesize \bf 37.3 &
\footnotesize \bf 5.4 &  \footnotesize \bf 20.8 &
\footnotesize \bf 5.5 &  \footnotesize \bf 21.9 &
\footnotesize \bf 5.8 &  \footnotesize \bf 37.3 &
\footnotesize \bf 5.8 &  \footnotesize \bf 39.1 &
\footnotesize \bf 6.4 &  \footnotesize \bf 40.5 &
\footnotesize \bf 5.9 &  \footnotesize \bf 42.9
\end{tabular}
\caption{Mean and maximal errors of one-shot DoEs in correlation predictions for analytical models.}
\label{tab:citl_fce}
\end{table*}
\begin{table*}[h!]
\centering
\vspace{6pt}
\tabcolsep=2pt
\begin{tabular}{cc|cc|cc|cc|cc|cc|cc|cc|cc}
& & \multicolumn{2}{c|} {\bfseries AE} & \multicolumn{2}{c|} {\bfseries EMM} & \multicolumn{2}{c|} {\bfseries ML$_2$} & \multicolumn{2}{c|} {\bfseries Dopt} & \multicolumn{2}{c|} {\bfseries PMCC} & \multicolumn{2}{c|} {\bfseries SRCC} & \multicolumn{2}{c|} {\bfseries KRCC} & \multicolumn{2}{c} {\bfseries CN}\\
&    points & \footnotesize mean & \footnotesize max & \footnotesize mean & \footnotesize max & \footnotesize mean & \footnotesize max & \footnotesize mean & \footnotesize max & \footnotesize mean & \footnotesize max & \footnotesize mean & \footnotesize max & \footnotesize mean & \footnotesize max & \footnotesize mean & \footnotesize max\\
\hline
\hline
\multirow{2}{*}{\rotatebox{90}{free}} & 10  &
\footnotesize 5.9 & \footnotesize 32.0 &
\footnotesize 6.2 & \footnotesize 29.4 &
\footnotesize 5.1 & \footnotesize 20.5 &
\footnotesize 4.5 & \footnotesize 33.6 &
\footnotesize 8.4 & \footnotesize 51.4 &
\footnotesize 7.3 & \footnotesize 48.5 &
\footnotesize 7.5 & \footnotesize 46.5 &
\footnotesize 8.6 & \footnotesize 59.9
\\
& 20  &
\footnotesize 3.3 & \footnotesize 16.5 &
\footnotesize 4.6 & \footnotesize 19.7 &
\footnotesize 3.1 & \footnotesize 14.3 &
\footnotesize 2.9 & \footnotesize 20.7 &
\footnotesize 5.1 & \footnotesize 34.2 &
\footnotesize 4.4 & \footnotesize 32.4 &
\footnotesize 4.4 & \footnotesize 29.9 &
\footnotesize 4.4 & \footnotesize 37.4
\\
& 30 &
\footnotesize 2.1 & \footnotesize 12.0 &
\footnotesize 4.3 & \footnotesize 17.1 &
\footnotesize 2.2 & \footnotesize 10.6 &
\footnotesize 3.0 & \footnotesize 17.4 &
\footnotesize 3.6 & \footnotesize 25.2 &
\footnotesize 3.3 & \footnotesize 22.7 &
\footnotesize 3.2 & \footnotesize 26.1 &
\footnotesize 2.9 & \footnotesize 22.0
\\
& 40 &
\footnotesize 1.5 & \footnotesize 10.0 &
\footnotesize 4.1 & \footnotesize 16.0 &
\footnotesize 1.7 & \footnotesize 8.5 &
\footnotesize 2.5 & \footnotesize 13.0 &
\footnotesize 2.6 & \footnotesize 19.0 &
\footnotesize 2.4 & \footnotesize 23.4 &
\footnotesize 2.5 & \footnotesize 19.6 &
\footnotesize 2.2 & \footnotesize 17.0
\\
\cline{2-18}
& overall&
\footnotesize \bf 3.2 & \footnotesize \bf 17.6 &
\footnotesize \bf 4.8 & \footnotesize \bf 20.6 &
\footnotesize \bf 3.0 & \footnotesize \bf 13.5 &
\footnotesize \bf 3.2 & \footnotesize \bf 21.2 &
\footnotesize \bf 4.9 & \footnotesize \bf 32.5 &
\footnotesize \bf 4.4 & \footnotesize \bf 31.8 &
\footnotesize \bf 4.4 & \footnotesize \bf 30.5 &
\footnotesize \bf 4.5 & \footnotesize \bf 34.1
\\
\hline
\hline
\multirow{2}{*}{\rotatebox{90}{LH}} & 10  &
\footnotesize 6.7 & \footnotesize 28.9 &
\footnotesize 9.8 & \footnotesize 31.1 &
\footnotesize 4.5 & \footnotesize 11.2 &
\footnotesize 6.8 & \footnotesize 20.3 &
\footnotesize 5.5 & \footnotesize 31.4 &
\footnotesize 5.8 & \footnotesize 28.4 &
\footnotesize 5.6 & \footnotesize 33.0 &
\footnotesize 5.3 & \footnotesize 30.7
\\
& 20  &
\footnotesize 4.8 & \footnotesize 15.4 &
\footnotesize 7.1 & \footnotesize 25.3 &
\footnotesize 2.8 & \footnotesize 10.2 &
\footnotesize 6.0 & \footnotesize 19.3 &
\footnotesize 3.1 & \footnotesize 20.8 &
\footnotesize 3.1 & \footnotesize 18.5 &
\footnotesize 3.0 & \footnotesize 18.6 &
\footnotesize 3.0 & \footnotesize 25.2
\\
& 30 &
\footnotesize 2.6 & \footnotesize 11.3 &
\footnotesize 5.3 & \footnotesize 22.9 &
\footnotesize 2.3 & \footnotesize 8.7 &
\footnotesize 2.5 & \footnotesize 9.6 &
\footnotesize 2.1 & \footnotesize 21.8 &
\footnotesize 2.2 & \footnotesize 18.2 &
\footnotesize 2.3 & \footnotesize 17.5 &
\footnotesize 2.2 & \footnotesize 20.5
\\
& 40 &
\footnotesize 2.5 & \footnotesize 9.0 &
\footnotesize 4.9 & \footnotesize 19.2 &
\footnotesize 1.6 & \footnotesize 5.9 &
\footnotesize 2.3 & \footnotesize 8.8 &
\footnotesize 1.6 & \footnotesize 11.3 &
\footnotesize 1.6 & \footnotesize 12.6 &
\footnotesize 1.8 & \footnotesize 15.5 &
\footnotesize 1.7 & \footnotesize 15.1
\\
\cline{2-18}
& overall&
\footnotesize \bf 4.2 & \footnotesize \bf 16.2 &
\footnotesize \bf 6.8 & \footnotesize \bf 24.6 &
\footnotesize \bf 2.8 & \footnotesize \bf 9.0 &
\footnotesize \bf 4.4 & \footnotesize \bf 14.5 &
\footnotesize \bf 3.1 & \footnotesize \bf 21.3 &
\footnotesize \bf 3.2 & \footnotesize \bf 19.4 &
\footnotesize \bf 3.2 & \footnotesize \bf 21.2 &
\footnotesize \bf 3.1 & \footnotesize \bf 22.9
\\
\end{tabular}
\caption{Mean and maximal errors of sequential DoEs in correlation predictions for analytical models.}
\label{tab:citl_fce_seq}
\end{table*}

The results on the sampling-based SA for analytical models can be
summarised as follows:
\begin{itemize}
\item The overall best results were achieved by the ML$_2$ optimal LH
  designs in terms of the average achieved errors as well as in their
  small variance. The results of the ML$_2$ optimal free designs were
  slightly worse and comparable to D-optimal designs. Free variant of
  both designs suffer from larger variance.

\item The AE and EMM criteria provided in average much better free
  designs comparing the their LH variant. However the variance is high
  for both of them. The quality of AE optimal free design can be
  classified as very good and comparable to results obtained by ML$_2$
  and Dopt criteria. On the other hand EMM criterion provided
  relatively bad designs, which were only very slowly improved by
  their sequential extension.

\item On the contrary, the orthogonality-based criteria achieved much
  better results for LH designs comparing to their free variant. All
  of these criteria provided comparable LH designs and can be
  evaluated as very good. However, their predictions suffer from very
  large variance.

\end{itemize}

\section{Sensitivity analysis on truss structures}
\label{sensitivity_kce}

The sensitivity analysis study presented in the previous section is aimed
on two-dimensional analytical problems. Therefore, this section is
devoted to illustrative engineering problems with higher dimensions. We
have chosen two models of truss structures commonly used as benchmarks
for sizing optimisation.

\begin{figure*}[h!]
\centering
\begin{minipage}[c]{6cm}
    {\includegraphics[width=5.3cm]{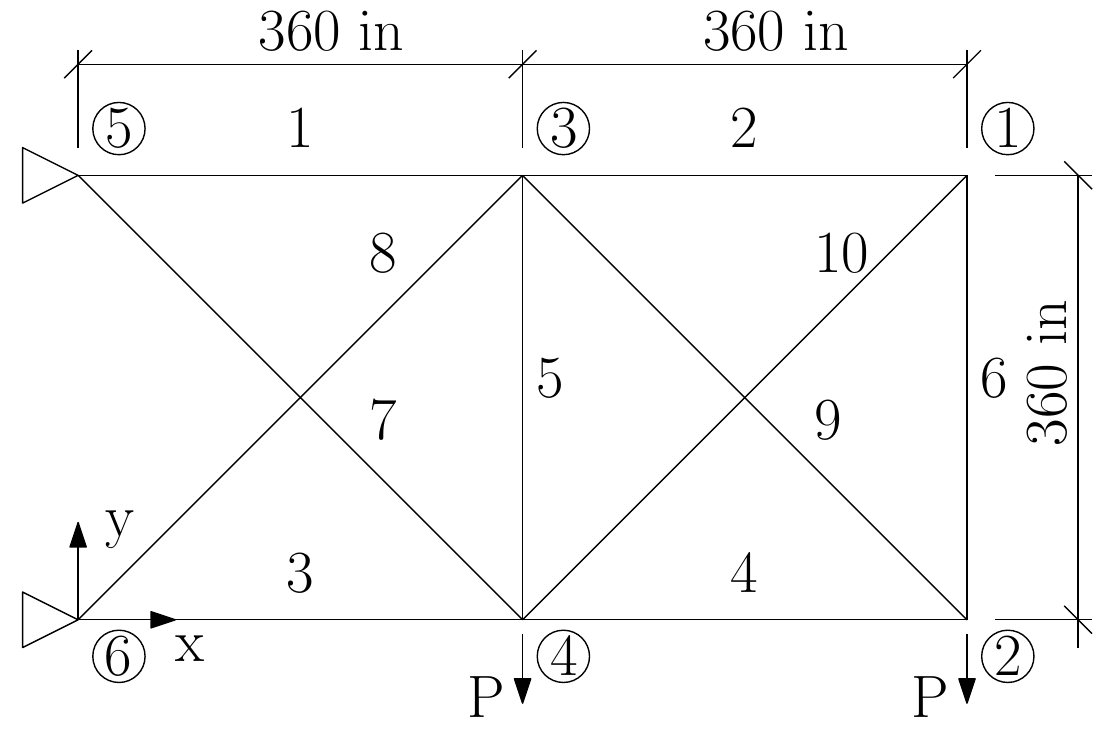}}
\end{minipage}
\begin{minipage}[c]{6cm}
    \begin{tabular}[p]{|l c|}
    \hline
    Material: & Aluminum\\
    Specific weight: & 0.1 lb/in$^3$\\
    Young's modulus: & $10^7$ psi\\
    Loading P: & 100 kips\\
    \hline
    \end{tabular}
\end{minipage}
\caption{Scheme of a ten-bar truss structure together with material
  parameters and values of applied loading.}
\label{fig:10bar}
\end{figure*}

The first one represents a ten-bar truss structure \cite{Venkaya:1971}
shown in Figure \ref{fig:10bar}. The design variables are the
cross-sectional areas of the bars. This benchmark is defined with two
types of variables: continuous and discrete one. Here we focus on a
discrete formulation with discrete values of cross-sectional areas
together with values of material properties and loading taken from
\cite{Lemonge:2003}. All ten cross-sectional areas have the same $42$
feasible discrete values (in$^2$): 1.62, 1.80, 1.99, 2.13, 2.38, 2.62,
2.63, 2.88, 2.83, 3.09, 3.13, 3.38, 3.47, 3.55, 3.63, 3.84, 3.87,
3.88, 4.18, 4.22, 4.49, 4.59, 4.80, 4.97, 5.12, 5.74, 7.22, 7.97,
11.50, 13.50, 13.90, 14.20, 15.50, 16.00, 16.90, 18.80, 19.90, 22.00,
22.90, 26.50, 30.00, 33.50.

\begin{figure*}[h!]
\centering
\begin{minipage}[c]{6cm}
    {\includegraphics[width=5.3cm]{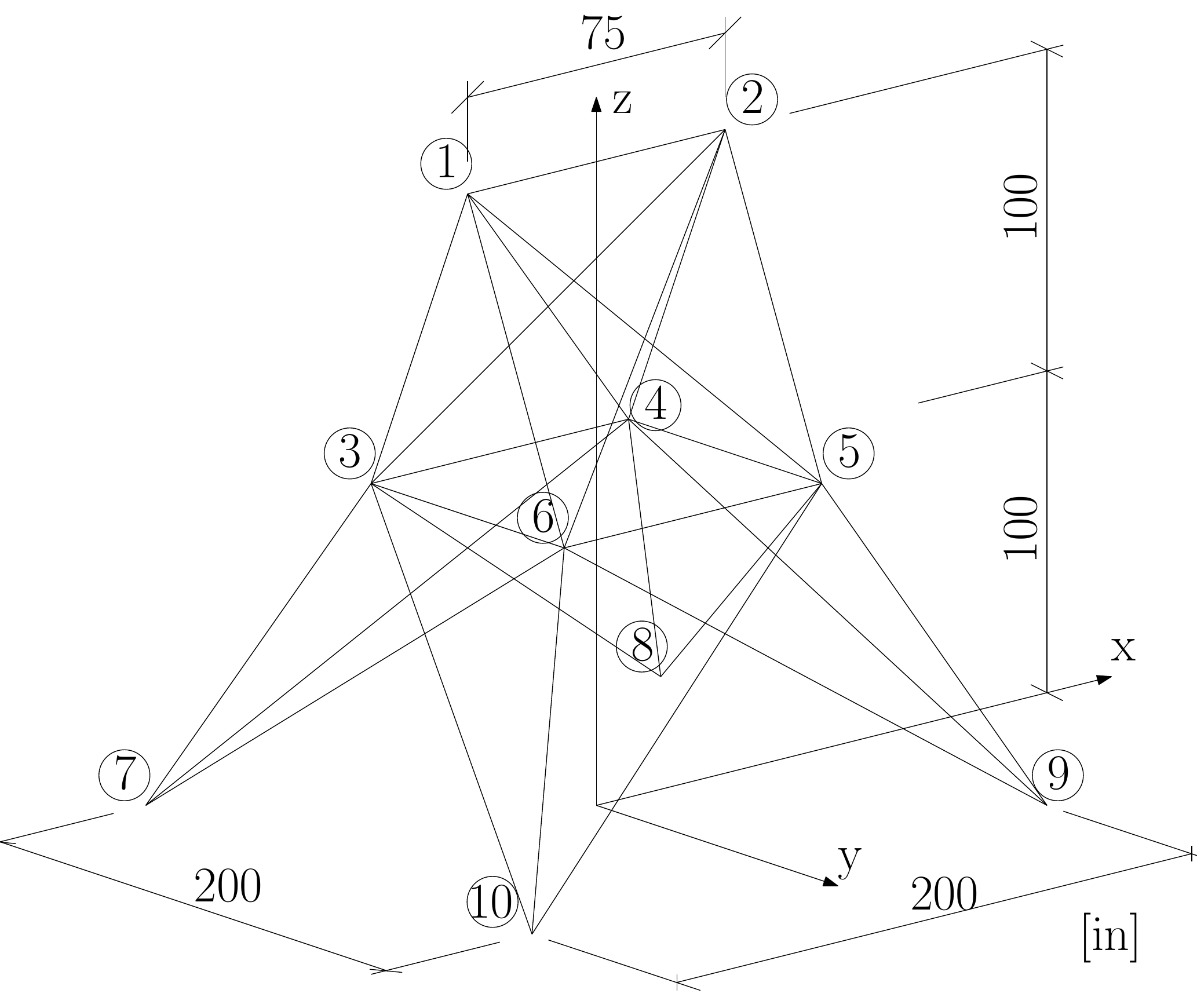}}
\end{minipage}
\begin{minipage}[c]{6cm}
    \centering
    \begin{tabular}[p]{|l c|}
    \hline
    Material: & Aluminum\\
    Specific weight: & 0,1 lb/in$^3$\\
    Young's modulus: & $10^7$ psi\\
    \hline
    \end{tabular}

\par

    \begin{tabular}[p]{l c c c}
      \hline
      \multicolumn{4}{c}{Nodal loads} \\
    \hline
    Node & F$_x$ & F$_y$ & F$_z$\\
    \hline
    1 & 1.0 & -10.0 & -10.0\\
    2 & 0.0 & -10.0 & -10.0\\
    3 & 0.5 &   0.0 &   0.0\\
    6 & 0.6 &   0.0 &   0.0\\
    \hline
    \end{tabular}
\end{minipage}
\caption{Scheme of a 25-bar truss structure together with material
  parameters and values of applied loading.}
\label{fig:25bar}
\end{figure*}

The second model concerns a 25-bar truss structure with a geometry,
material properties and loading given in Figure \ref{fig:25bar}.
Thanks to the symmetry of the structure, the bars can be organised into
eight groups. The bars in one group have the same
cross-sectional areas and hence, there are only eight design variables.
These variables are again discrete with $30$ feasible values (in$^2$):
0.1, 0.2, 0.3, 0.4, 0.5, 0.6, 0.7, 0.8, 0.9, 1.0, 1.1, 1.2, 1.3, 1.4,
1.5, 1.6, 1.7, 1.8, 1.9, 2.0, 2.1, 2.2, 2.3, 2.4, 2.5, 2.6, 2.8, 3.0,
3.2, 3.4, see \cite{Wu:1995}.

The response of these models consists of three components: total
weight of the structure $w$, maximal deflection $d$ and maximal stress
$s$. Because of higher dimensions of these problems, we have generated
the optimal DoEs only with LH restriction, since they can be optimised
more easily. This restriction automatically specifies the number of
corresponding design points, which has to be equal to the number of
feasible values ($42$ for the ten-bar truss and $30$ for the 25-bar
truss) or their multiples in case of sequential designs.

\begin{figure*}[h!]
\centering
\includegraphics{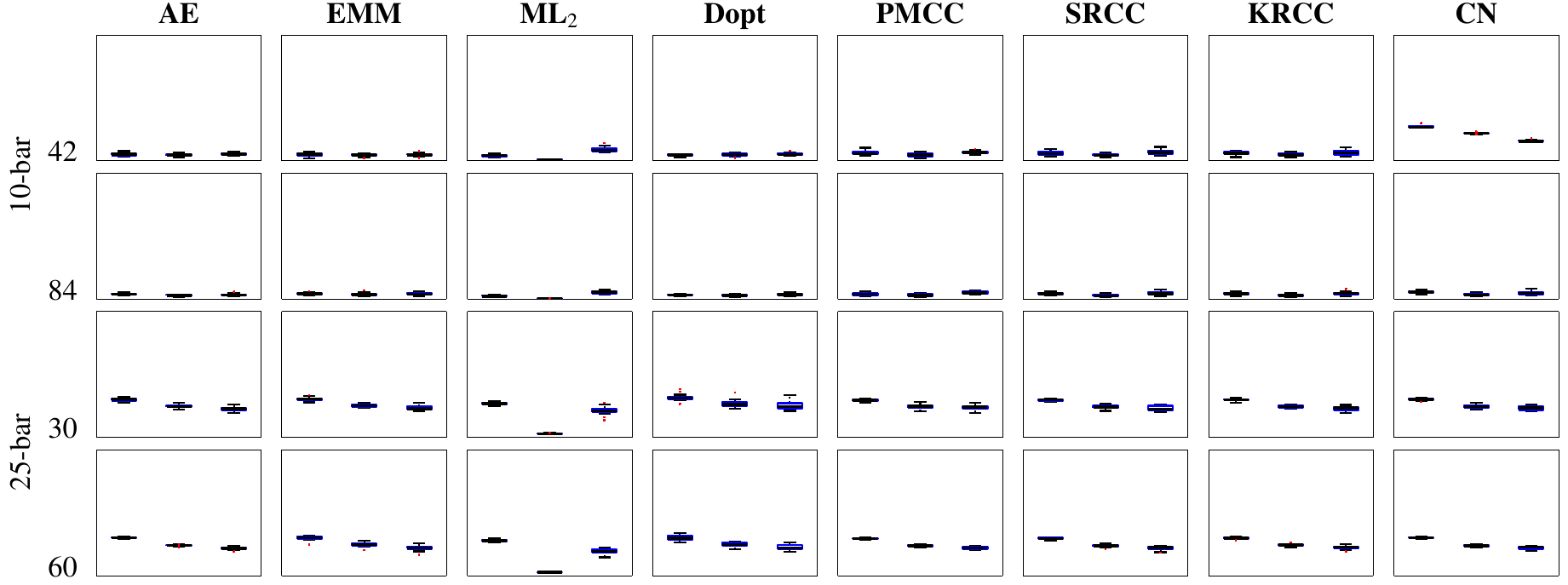}
\caption{Statistics on results in estimation of
    parameter-response correlations for structural models. Each
    rectangle contains three box plots representing the distribution
    of the errors in correlation prediction for particular model
    responses: the total weight $w$, the maximal deflection $d$ and
    the maximal stress $s$. The~scale is $0$ at the bottom and $1$ at
    the top per rectangle. Each rectangle refers to optimal designs
    according to one criterion associated with the corresponding
    column and to the chosen structure and number of design points
    associated with~the corresponding row.}
\label{fig:citl_kce}
\end{figure*}

A simulated annealing method with the same parameters as described in
Section \ref{mutual} but with a maximum of $10^7$ iterations was
employed to generate $20$ optimal DoEs using each criterion under the
study.  The parameter-response correlations were then estimated using
the obtained optimal DoEs and compared with the correlations computed
using the {\bf Large} DoEs consisting of $2 \cdot 10^7$ samples
generated by the Monte Carlo method. The statistics on the errors
$\epsilon$ in estimation of parameter-response correlations is
demonstrated again in terms of box plots independently for each
criterion and each model response component, see Figure
\ref{fig:citl_kce}.

For a clearer summary of the obtained results, the mean and
maximal errors in correlation predictions multiplied again by $100$
are listed in Table \ref{tab:citl_kce}.
\renewcommand{\tabcolsep}{0.5mm}
\begin{table*}[h!]
\centering
\begin{tabular}{ccc|cc|cc|cc|cc|cc|cc|cc|cc}
\multicolumn{3}{c|}{\bfseries  Model} & \multicolumn{2}{c|}{\bfseries  AE} & \multicolumn{2}{c|}{\bfseries EMM} & \multicolumn{2}{c|} {\bfseries ML$_2$} & \multicolumn{2}{c|} {\bfseries Dopt} & \multicolumn{2}{c|} {\bfseries PMCC} & \multicolumn{2}{c|} {\bfseries SRCC} & \multicolumn{2}{c|} {\bfseries KRCC} & \multicolumn{2}{c} {\bfseries CN}\\
&  &        & \footnotesize mean & \footnotesize max & \footnotesize mean & \footnotesize max & \footnotesize mean & \footnotesize max & \footnotesize mean & \footnotesize max & \footnotesize mean & \footnotesize max & \footnotesize mean & \footnotesize max & \footnotesize mean & \footnotesize max & \footnotesize mean & \footnotesize max\\
\hline
\hline
\multirow{8}{*}{\rotatebox{90}{10-bar}}  & \multirow{4}{*}{\rotatebox{90}{42}} & $w$
& \footnotesize 5.1 & \footnotesize 7.5 & \footnotesize 4.9 & \footnotesize 7.5 & \footnotesize 4.1 & \footnotesize 5.5 & \footnotesize 4.3 & \footnotesize 5.4 & \footnotesize 6.2 & \footnotesize 10.3 & \footnotesize 5.9 & \footnotesize 9.2 & \footnotesize 5.8 & \footnotesize 7.5 & \footnotesize 27.0 & \footnotesize 29.7  \\
& & $d$ & \footnotesize 4.5 & \footnotesize 6.7 & \footnotesize 4.4 & \footnotesize 5.6 & \footnotesize 0.7 & \footnotesize 1.1 & \footnotesize 4.6 & \footnotesize 6.5 & \footnotesize 4.5 & \footnotesize 7.3 & \footnotesize 4.3 & \footnotesize 6.2 & \footnotesize 4.7 & \footnotesize 7.1 & \footnotesize 21.7 & \footnotesize 23.0  \\
& & $s$ & \footnotesize 5.1 & \footnotesize 7.3 & \footnotesize 4.7 & \footnotesize 7.7 & \footnotesize 8.9 & \footnotesize 13.9 & \footnotesize 5.1 & \footnotesize 8.0 & \footnotesize 6.5 & \footnotesize 9.2 & \footnotesize 6.8 & \footnotesize 11.0 & \footnotesize 6.6 & \footnotesize 10.7 & \footnotesize 15.7 & \footnotesize 17.7 \\
\cline{3-19}
& & overall
& \footnotesize \bf 4.9 & \footnotesize \bf	7.5 & \footnotesize \bf	4.7 & \footnotesize \bf	 7.7 & \footnotesize \bf	4.6 & \footnotesize \bf	13.9 & \footnotesize \bf	4.7 & \footnotesize \bf	8.0 & \footnotesize \bf	5.7 & \footnotesize \bf	10.3 & \footnotesize \bf	 5.7 & \footnotesize \bf	11.0 & \footnotesize \bf	5.7 & \footnotesize \bf	10.7 & \footnotesize \bf	21.5 & \footnotesize \bf	29.7 \\
\cline{2-19}
& \multirow{4}{*}{\rotatebox{90}{84}} & $w$
& \footnotesize 3.9 & \footnotesize 5.1 & \footnotesize 4.1 & \footnotesize 6.1 & \footnotesize 2.3 & \footnotesize 3.6 & \footnotesize 3.0 & \footnotesize 4.2 & \footnotesize 3.7 & \footnotesize 6.3 & \footnotesize 3.9 & \footnotesize 6.1 & \footnotesize 4.1 & \footnotesize 5.8 & \footnotesize 5.4 & \footnotesize 7.7  \\
& & $d$ & \footnotesize 2.7 & \footnotesize 3.6 & \footnotesize 3.6 & \footnotesize 6.7 & \footnotesize 0.4 & \footnotesize 0.7 & \footnotesize 2.9 & \footnotesize 4.0 & \footnotesize 3.1 & \footnotesize 5.0 & \footnotesize 2.7 & \footnotesize 4.6 & \footnotesize 2.9 & \footnotesize 4.4 & \footnotesize 3.4 & \footnotesize 5.5  \\
& & $s$ & \footnotesize 3.3 & \footnotesize 5.7 & \footnotesize 3.9 & \footnotesize 6.0 & \footnotesize 5.4 & \footnotesize 7.7 & \footnotesize 3.7 & \footnotesize 5.8 & \footnotesize 5.0 & \footnotesize 6.7 & \footnotesize 4.3 & \footnotesize 7.7 & \footnotesize 4.4 & \footnotesize 8.0 & \footnotesize 4.8 & \footnotesize 8.0 \\
\cline{3-19}
& & overall
& \footnotesize \bf 3.3 & \footnotesize \bf	5.7 & \footnotesize \bf	3.9 & \footnotesize \bf	 6.7 & \footnotesize \bf	2.7 & \footnotesize \bf	7.7 & \footnotesize \bf	3.2 & \footnotesize \bf	5.8 & \footnotesize \bf	3.9 & \footnotesize \bf	6.7 & \footnotesize \bf	3.6 & \footnotesize \bf	7.7 & \footnotesize \bf	3.8 & \footnotesize \bf	8.0 & \footnotesize \bf	 4.5 & \footnotesize \bf	8.0 \\
\hline
\hline
\multirow{8}{*}{\rotatebox{90}{25-bar}} & \multirow{4}{*}{\rotatebox{90}{30}} & $w$
& \footnotesize 30.0 & \footnotesize 32.3 & \footnotesize 30.4 & \footnotesize 33.5 & \footnotesize 26.8 & \footnotesize 28.9 & \footnotesize 31.6 & \footnotesize 38.3 & \footnotesize 29.5 & \footnotesize 30.8 & \footnotesize 29.6 & \footnotesize 30.7 & \footnotesize 29.9 & \footnotesize 31.4 & \footnotesize 30.0 & \footnotesize 31.2  \\
& & $d$ & \footnotesize 24.6 & \footnotesize 27.2 & \footnotesize 25.1 & \footnotesize 27.5 & \footnotesize 23.6 & \footnotesize 28.3 & \footnotesize 26.8 & \footnotesize 35.7 & \footnotesize 24.4 & \footnotesize 28.1 & \footnotesize 24.1 & \footnotesize 27.1 & \footnotesize 24.3 & \footnotesize 25.8 & \footnotesize 24.5 & \footnotesize 27.2  \\
& & $s$ & \footnotesize 22.6 & \footnotesize 25.9 & \footnotesize 23.2 & \footnotesize 27.6 & \footnotesize 21.2 & \footnotesize 27.4 & \footnotesize 25.2 & \footnotesize 33.4 & \footnotesize 23.8 & \footnotesize 27.2 & \footnotesize 23.0 & \footnotesize 26.0 & \footnotesize 22.7 & \footnotesize 25.8 & \footnotesize 23.4 & \footnotesize 26.3  \\
\cline{3-19}
& & overall & \footnotesize \bf
25.7 & \footnotesize \bf	32.3 & \footnotesize \bf	26.2 & \footnotesize \bf	33.5 & \footnotesize \bf	23.9 & \footnotesize \bf	28.9 & \footnotesize \bf	27.9 & \footnotesize \bf	38.3 & \footnotesize \bf	25.9 & \footnotesize \bf	30.8 & \footnotesize \bf	25.6 & \footnotesize \bf	 30.7 & \footnotesize \bf	25.6 & \footnotesize \bf	31.4 & \footnotesize \bf	 26.0 & \footnotesize \bf	31.2\\
\cline{2-19}
& \multirow{4}{*}{\rotatebox{90}{60}} & $w$
& \footnotesize 30.0 & \footnotesize 30.9 & \footnotesize 30.0 & \footnotesize 31.9 & \footnotesize 27.8 & \footnotesize 29.5 & \footnotesize 30.2 & \footnotesize 34.1 & \footnotesize 29.4 & \footnotesize 30.4 & \footnotesize 29.4 & \footnotesize 30.4 & \footnotesize 29.9 & \footnotesize 31.4 & \footnotesize 29.9 & \footnotesize 30.9  \\
& & $d$ & \footnotesize 23.9 & \footnotesize 25.1 & \footnotesize 24.7 & \footnotesize 27.7 & \footnotesize 22.5 & \footnotesize 25.5 & \footnotesize 24.9 & \footnotesize 27.3 & \footnotesize 23.8 & \footnotesize 25.3 & \footnotesize 23.6 & \footnotesize 25.6 & \footnotesize 24.2 & \footnotesize 26.5 & \footnotesize 23.8 & \footnotesize 25.1  \\
& & $s$ & \footnotesize 21.6 & \footnotesize 23.7 & \footnotesize 22.3 & \footnotesize 25.8 & \footnotesize 19.3 & \footnotesize 22.6 & \footnotesize 22.6 & \footnotesize 26.4 & \footnotesize 22.1 & \footnotesize 23.9 & \footnotesize 21.9 & \footnotesize 23.9 & \footnotesize 22.5 & \footnotesize 25.1 & \footnotesize 22.2 & \footnotesize 23.8  \\
\cline{3-19}
& & overall & \footnotesize \bf
25.2 & \footnotesize \bf	30.9 & \footnotesize \bf	25.7 & \footnotesize \bf	31.9 & \footnotesize \bf	23.2 & \footnotesize \bf	29.5 & \footnotesize \bf	25.9 & \footnotesize \bf    34.1 & \footnotesize \bf	25.1 & \footnotesize \bf	30.4 & \footnotesize \bf	25.0 & \footnotesize \bf	30.4 & \footnotesize \bf	 25.5 & \footnotesize \bf	31.4 & \footnotesize \bf	25.3 & \footnotesize \bf	 30.9 \\
\end{tabular}
\caption{Mean and maximal errors in correlation predictions for structural models.}
\label{tab:citl_kce}
\end{table*}

To conclude the obtained results on the sampling-based SA for
structural models we can formulate following points:
\begin{itemize}
\item Apart from the CN optimal designs, all the other designs
  achieved very good results in predicting sensitivities for the
  ten-bar truss structure.  The CN designs achieved surprisingly bad
  results in predicting sensitivities to all three response
  components. Nevertheless, these results were significantly improved
  by extension of the design.

\item All space-filling criteria achieved better results for the
  ten-bar truss structure than all orthogonality based criteria. The
  smallest mean error in prediction was obtained by the ML$_2$
  designs, however it is accompanied by relatively big differences
  among errors corresponding to particular response components. The
  AE, EMM and Dopt designs provided more balanced results.

\item We are not able to find differences among the studied criteria
  based on the results for the $25$-bar structure. Obviously, the
  model is too complex to estimate the parameter-response
  sensitivities using the design with only $30$ or $60$ points
  in the eight-dimensional design space and hence, all the criteria
  led to comparably bad predictions.

\end{itemize}

\section{Conclusions}
\label{concl}

This paper reviews eight criteria used for optimizing a design of
experiments and presents their comparison in terms of ease of their
optimisation, their mutual qualities and their suitability for usage
in sampling-based SA on $15$ analytical and two structural models.
The overall results can be summarised in several following
conclusions:
\begin{itemize}
\item CN criterion was poorly evaluated in terms of all presented
  aspects. It may make the optimisation process more complicated by
  its tendency to the higher number of local extremes. The resulting
  designs have very poor space-filling properties and the subsequent
  sensitivity predictions for analytical as well as for structural
  models contain large errors.

\item The correlation-based criteria (PMCC, SRCC and KRCC) may also
  pose some difficulties during the optimisation process: SRCC and
  KRCC due to their discrete nature and PMCC due to its stronger
  tendency to the multi-modality. All these criteria provide the
  designs with very bad space-filling property and even LH restriction
  does not improve it significantly. While the free designs achieved
  also very bad results in SA, the predictions of LH designs can be
  evaluated as very good, but they suffer from higher variances.

\item AE and EMM criteria do not exhibit any explicit difficulties
  regarding the optimisation process. The advantage of these criteria
  is that their interpretation is very simple and computation very
  fast. They provide designs with similar properties: good
  space-filling and a moderate level of orthogonality. The free
  designs achieved good results in sensitivity predictions but also
  with a high variance among the results. The results of EMM criterion
  are in overall worse then those of AE criterion. Their qualities are
  generally deteriorated by applying the LH restriction.

\item Finally, the best results in sensitivity predictions were
  obtained using the LH designs optimised w.r.t. the ML$_2$ criterion.
  The D-optimal designs were slightly worse and comparable to AE free
  designs or LH designs optimised w.r.t. the correlation-based
  criteria. ML$_2$ and Dopt criteria provide designs with moderate
  space-filling properties, but they have a good level of
  orthogonality.  The LH restriction slightly improves the results of
  both.  The ML$_2$ designs achieved less varying results in the SA
  for analytical functions, while the D-optimal designs provided more
  balanced results in the SA for the ten-bar truss structure. The
  ML$_2$ designs also significantly overcame all the other designs in
  terms of the projective properties. An important shortcoming of the
  D-optimal criterion concerns its formulation and optimisation.
  Besides the fact that the criterion results in an excessive number
  of local extremes, the principal drawback lies in the necessity of
  its Bayesian modification for elimination of duplicates or closely
  neighbouring points. Therefore, the ML$_2$ criterion, which is not
  so common in DoE optimisation, can be considered as a winner in the
  presented competitive comparisons.

\end{itemize}

\section*{Acknowledgment}
The financial support of this work by the Czech Science Foundation
(project Nos. 105/11/P370 and 105/12/1146) is gratefully acknowledged.

\bibliographystyle{elsarticle-num}
\bibliography{liter}

\end{document}